\pdfoutput=1
\documentclass{article}

\usepackage{PRIMEarxiv}

\usepackage[utf8]{inputenc}
\usepackage[T1]{fontenc}
\usepackage{amssymb}
\usepackage{amsmath}
\usepackage{amsfonts}
\usepackage{booktabs}
\usepackage{longtable}
\usepackage{multirow}
\usepackage{array}
\usepackage{tabularx}
\usepackage{float}
\usepackage{placeins}
\usepackage{nicefrac}
\usepackage{microtype}
\usepackage{graphicx}
\graphicspath{{media/}}
\usepackage{xcolor}
\usepackage{fancyhdr}
\usepackage[numbers,sort&compress]{natbib}
\usepackage{url}
\usepackage{hyperref}
\hypersetup{colorlinks=true, linkcolor=blue, citecolor=blue, urlcolor=blue}
\hypersetup{pdftitle={Beauty Is Human: When People Love AI-Generated Art, They Think a Human Made It},
  pdfauthor={Zihan Chen, Di Zhu, Lei Nico Zheng},
  pdfkeywords={AI-generated art, perceived authorship, aesthetic preference, algorithm aversion, generative AI, human-AI interaction, AI disclosure},
  pdfsubject={Liking and perceived human authorship of AI-generated art and poetry}}

\newcommand{\AIAndHAll}{38.1\%}
\newcommand{\AccMean}{60.6\%}
\newcommand{\AccMedian}{60\%}
\newcommand{\AttAIAndHFive}{33.5\%}
\newcommand{\AttAIAndHOne}{37.3\%}
\newcommand{\AttAIBelHFive}{56\%}
\newcommand{\AttAIBelHOne}{80\%}
\newcommand{\AttAIOthersHFive}{38\%}
\newcommand{\AttAIOthersHOne}{54\%}
\newcommand{\AttAccAIFavFive}{44\%}
\newcommand{\AttAccAIFavOne}{20\%}
\newcommand{\AttAccFive}{60.3\%}
\newcommand{\AttAccOne}{63.9\%}
\newcommand{\AttBelHIsAIOne}{43\%}
\newcommand{\AttCFive}{$-$0.05}
\newcommand{\AttCOne}{$-$0.08}
\newcommand{\AttCalledHFive}{64\%}
\newcommand{\AttCalledHOne}{88\%}
\newcommand{\AttDFive}{0.53}
\newcommand{\AttDOne}{0.72}
\newcommand{\AttFavAIFive}{60\%}
\newcommand{\AttFavAIOne}{47\%}
\newcommand{\AttFiveContrast}{8.1}
\newcommand{\AttFiveContrastHi}{11.1}
\newcommand{\AttFiveContrastLo}{5.1}
\newcommand{\AttRawPicSlope}{6.3}
\newcommand{\AttSamePicSlope}{2.2}
\newcommand{\AttSamePicSlopeSE}{0.5}
\newcommand{\AttSlope}{2.0}
\newcommand{\AttSlopeHi}{2.6}
\newcommand{\AttSlopeLo}{1.4}
\newcommand{\AttTwfeFive}{16.6}
\newcommand{\AttTwfeOne}{24.8}
\newcommand{\AttTwfeSEFive}{1.2}
\newcommand{\ConfContrast}{4.2}
\newcommand{\ConfContrastHi}{6.6}
\newcommand{\ConfContrastLo}{1.8}
\newcommand{\ConfUns}{4.2}
\newcommand{\ConfUnsAI}{0.6}
\newcommand{\ConfUnsAISE}{1.1}
\newcommand{\ConfUnsH}{4.8}
\newcommand{\ConfUnsHSE}{0.4}
\newcommand{\ConfUnsSE}{0.4}
\newcommand{\CorrAIAbs}{21.3}
\newcommand{\CorrH}{20.8}
\newcommand{\DeltaC}{$-$0.66}
\newcommand{\DeltaCHi}{$-$0.63}
\newcommand{\DeltaCLo}{$-$0.68}
\newcommand{\DeltaD}{$+$0.16}
\newcommand{\DeltaDHi}{$+$0.22}
\newcommand{\DeltaDLo}{$+$0.11}
\newcommand{\DeltaHmF}{$-$0.2}
\newcommand{\DeltaHmFHi}{$+$1.3}
\newcommand{\DeltaHmFLo}{$-$1.6}
\newcommand{\ErrGoodContrast}{$-$0.4}
\newcommand{\ErrGoodHi}{$+$2.0}
\newcommand{\ErrGoodLo}{$-$2.7}
\newcommand{\ExAGFans}{22\%}
\newcommand{\ExAGOthers}{9\%}
\newcommand{\ExPSFans}{96\%}
\newcommand{\ExPSOthers}{83\%}
\newcommand{\FamAIContrast}{14.7}
\newcommand{\FamAIContrastHi}{24.1}
\newcommand{\FamAIContrastLo}{5.4}
\newcommand{\FamAIContrastP}{.002}
\newcommand{\FamAIExpert}{7.4}
\newcommand{\FamAIExpertN}{63}
\newcommand{\FamAIExpertSE}{4.8}
\newcommand{\FamAIRegular}{18.0}
\newcommand{\FamAIRegularSE}{1.2}
\newcommand{\FamHContrast}{$+$0.8}
\newcommand{\FamHContrastHi}{$+$4.0}
\newcommand{\FamHContrastLo}{$-$2.4}
\newcommand{\FamHExpert}{22.8}
\newcommand{\FamHExpertAcc}{66.9\%}
\newcommand{\FamHExpertN}{379}
\newcommand{\FamHExpertSE}{1.6}
\newcommand{\FamHZeroAcc}{59.0\%}
\newcommand{\FavAIOthers}{52.2\%}
\newcommand{\FavAIPaired}{3.5}
\newcommand{\FavAIPairedHi}{4.5}
\newcommand{\FavAIPairedLo}{2.4}
\newcommand{\FavAIShare}{55.6\%}
\newcommand{\FavCalledH}{76.6\%}
\newcommand{\FavCalledHAI}{68\%}
\newcommand{\FavCalledHH}{87\%}
\newcommand{\FavEqMca}{1.4\%}
\newcommand{\FavEqMcaChance}{1.8\%}
\newcommand{\FavEqMch}{13.0\%}
\newcommand{\FavEqMchChance}{2.3\%}
\newcommand{\FavEqMchRatio}{5.6}
\newcommand{\FlowAccEx}{59.5\%}
\newcommand{\FlowAccIn}{60.9\%}
\newcommand{\FlowAnsEx}{44.8}
\newcommand{\FlowAnsIn}{49.6}
\newcommand{\FlowExcl}{1{,}971}
\newcommand{\FlowExclPct}{17.7\%}
\newcommand{\FlowNoFav}{1{,}951}
\newcommand{\FlowNotJudged}{20}
\newcommand{\FlowOpMissEx}{42\%}
\newcommand{\FlowOpMissIn}{1\%}
\newcommand{\HeldBase}{21.8}
\newcommand{\HeldFour}{19.9}
\newcommand{\HeldN}{4{,}496}
\newcommand{\HumanRateAI}{42\%}
\newcommand{\HumanRateH}{63\%}
\newcommand{\ImgNGe}{49}
\newcommand{\JointMax}{40.8\%}
\newcommand{\JointMin}{33.5\%}
\newcommand{\LogitAME}{23.6}
\newcommand{\LogitOR}{3.87}
\newcommand{\LooMax}{22.5}
\newcommand{\LooMin}{21.5}
\newcommand{\Main}{22.1}
\newcommand{\MainAI}{21.8}
\newcommand{\MainAISE}{0.6}
\newcommand{\MainH}{22.4}
\newcommand{\MainHSE}{0.5}
\newcommand{\MainHi}{22.9}
\newcommand{\MainHiTwo}{24.6}
\newcommand{\MainLo}{21.3}
\newcommand{\MainLoTwo}{19.6}
\newcommand{\MainSE}{0.4}
\newcommand{\MchShareAI}{30\%}
\newcommand{\MchTopAIImg}{Paris Scene}
\newcommand{\MchTopAIN}{524}
\newcommand{\MchTopAIRank}{third most}
\newcommand{\NAIImg}{26}
\newcommand{\NFav}{9{,}152}
\newcommand{\NHumanImg}{24}
\newcommand{\NJudg}{542{,}208}
\newcommand{\NOpinion}{10{,}212}
\newcommand{\NPeople}{11{,}123}
\newcommand{\NRows}{453{,}865}
\newcommand{\NonfanSame}{54.6\%}
\newcommand{\NonfanSameAI}{46\%}
\newcommand{\NonfanSameH}{65\%}
\newcommand{\OpNeg}{33\%}
\newcommand{\OpNeu}{24\%}
\newcommand{\OpPos}{42\%}
\newcommand{\OverlapPics}{15}
\newcommand{\OverlapShareFive}{79\%}
\newcommand{\OverlapShareOne}{80\%}
\newcommand{\PAcc}{47.3\%}
\newcommand{\PBgWrite}{13.2}
\newcommand{\PBgWriteSE}{1.5}
\newcommand{\PGap}{0.89}
\newcommand{\PGapComp}{0.69}
\newcommand{\PGapMinusLabel}{0.08}
\newcommand{\PGapMinusLabelHi}{0.31}
\newcommand{\PGapMinusLabelLo}{$-$0.16}
\newcommand{\PGapSE}{0.08}
\newcommand{\PGoodAI}{5.30}
\newcommand{\PGoodH}{4.18}
\newcommand{\PHumanRateAI}{61\%}
\newcommand{\PHumanRateH}{55\%}
\newcommand{\PLabel}{0.81}
\newcommand{\PLabelComp}{0.64}
\newcommand{\PLabelSE}{0.09}
\newcommand{\PLooSlopeMax}{11.7}
\newcommand{\PLooSlopeMin}{11.2}
\newcommand{\PNAll}{696}
\newcommand{\PNJudg}{2{,}300}
\newcommand{\PNLabel}{466}
\newcommand{\PNPeople}{230}
\newcommand{\PNonTopCalledH}{52\%}
\newcommand{\PQualMax}{18.7}
\newcommand{\PQualMin}{11.9}
\newcommand{\PSlope}{11.5}
\newcommand{\PSlopeAI}{11.7}
\newcommand{\PSlopeContrast}{$-$0.3}
\newcommand{\PSlopeContrastHi}{$+$2.7}
\newcommand{\PSlopeContrastLo}{$-$3.3}
\newcommand{\PSlopeH}{11.4}
\newcommand{\PSlopeSE}{0.9}
\newcommand{\PTop}{23.6}
\newcommand{\PTopAI}{18.5}
\newcommand{\PTopCalledH}{73\%}
\newcommand{\PTopContrast}{14.3}
\newcommand{\PTopContrastHi}{24.1}
\newcommand{\PTopContrastLo}{4.4}
\newcommand{\PTopH}{32.7}
\newcommand{\PTopSE}{2.8}
\newcommand{\PTopUnique}{24.6}
\newcommand{\PTopUniqueN}{77}
\newcommand{\PTopUniqueSE}{5.0}
\newcommand{\PicHolm}{45}
\newcommand{\PicMax}{38.1}
\newcommand{\PicMin}{5.3}
\newcommand{\PicMu}{22.4}
\newcommand{\PicMuAI}{23.0}
\newcommand{\PicMuH}{21.9}
\newcommand{\PicMuSE}{1.2}
\newcommand{\PicTau}{7.4}
\newcommand{\RawPicGap}{24.4}
\newcommand{\RobNoMch}{20.7}
\newcommand{\RobSeen}{22.6}
\newcommand{\RobSeenN}{4{,}704}
\newcommand{\SamePicGap}{8.9}
\newcommand{\SamePicGapHi}{13.7}
\newcommand{\SamePicGapLo}{4.2}
\newcommand{\ShareImgAI}{52\%}
\newcommand{\SharedGap}{13.1}
\newcommand{\SharedGapHi}{18.2}
\newcommand{\SharedGapLo}{8.0}
\newcommand{\SharedGapN}{1{,}020}
\newcommand{\SlopeFavAI}{3.6}
\newcommand{\SlopeFavAISE}{0.4}
\newcommand{\SpillFour}{2.9}
\newcommand{\SpillRatio}{8}
\newcommand{\SrcContrast}{0.5}
\newcommand{\SrcContrastHi}{$+$2.1}
\newcommand{\SrcContrastHiTwo}{$+$5.6}
\newcommand{\SrcContrastLo}{$-$1.0}
\newcommand{\SrcContrastLoTwo}{$-$4.5}
\newcommand{\StratErrors}{23.4}
\newcommand{\StratErrorsN}{3{,}703}
\newcommand{\StratErrorsSE}{0.6}
\newcommand{\StratGoodH}{23.8}
\newcommand{\StratGoodHN}{1{,}209}
\newcommand{\StratGoodHSE}{1.0}
\newcommand{\StratImprAI}{17.1}
\newcommand{\StratMax}{25.1}
\newcommand{\StratMin}{17.1}
\newcommand{\StratWaldP}{$p < .001$}
\newcommand{\StyleEight}{19.0}
\newcommand{\StyleFour}{20.0}
\newcommand{\TFansMerit}{7.7\%}
\newcommand{\TImgFlaw}{33\%}
\newcommand{\TImgFlawShare}{13.9\%}
\newcommand{\TImgIntent}{68\%}
\newcommand{\TImgMerit}{80\%}
\newcommand{\TImgMeritShare}{1.7\%}
\newcommand{\TImgNoMerit}{22\%}
\newcommand{\TImgPolish}{28\%}
\newcommand{\TNImg}{49{,}268}
\newcommand{\TNImgPeople}{8{,}862}
\newcommand{\TNPoem}{5{,}238}
\newcommand{\TNPoemPeople}{950}
\newcommand{\TOthersMerit}{1.6\%}
\newcommand{\TPoemMerit}{90\%}
\newcommand{\TPoemMeritShare}{8.5\%}
\newcommand{\TPoemNoMerit}{7\%}
\newcommand{\TopImg}{Riverside Cafe}
\newcommand{\TopImgFavN}{642}
\newcommand{\TopImgHumanRate}{69\%}
\newcommand{\TopTenAI}{6}

\title{Beauty Is Human:\\When People Love AI-Generated Art,\\They Think a Human Made It}

\author{
  Zihan Chen \\
  Stevens Institute of Technology \\
  Hoboken, NJ, USA \\
  \texttt{zchen61@stevens.edu} \\
  \And
  Di Zhu \\
  Stevens Institute of Technology \\
  Hoboken, NJ, USA \\
  \texttt{dzhu1@stevens.edu} \\
  \And
  Lei Nico Zheng \\
  University of Massachusetts Boston \\
  Boston, MA, USA \\
  \texttt{lei.zheng@umb.edu} \\
}

\begin{document}
\maketitle

\begin{abstract}
Generative AI now produces images and poems that people cannot reliably distinguish from human-made work, so people often judge for themselves who made a work. We examine how these perceived-authorship judgments relate to aesthetic preference. In a public online AI Art Turing Test, \NPeople{} people judged whether each of 50 pictures (\NAIImg{} AI-generated) was made by a human and were asked for their favourite. Among the \NFav{} respondents who judged their favourite, a picture was \Main{} percentage points more likely to be judged human-made when it was the respondent's favourite, holding respondent and picture fixed. The association was similar for AI-generated and human-made favourites, was positive for all 50 pictures, and appeared under every reported detection strategy. Respondents most negative about AI art chose fewer AI-generated favourites (\AttFavAIOne{} against \AttFavAIFive{} among the most positive) but more often believed them human-made, partly because others also judged those pictures human. As a result, in every opinion group a third to two fifths of respondents named an AI-generated favourite they believed human-made. A reanalysis of a published experiment on AI-generated poetry, in which \PNPeople{} people rated each poem before judging its authorship, found the same pattern, although the AI poems were rated higher. When provenance is hidden, liking and perceived human authorship travel together, consistent with aesthetic response serving as evidence of a human maker and with beliefs about the maker shaping enjoyment. The findings qualify research on bias against AI art, which typically uses explicit labels, and inform AI disclosure policy.
\end{abstract}


\section{Introduction}
\label{sec:intro}

In late 2024, the writer of a widely read blog posted 50 pictures and asked readers to say which ones a human had made \citep{alexander2024results}. Twenty-four were well-regarded human works, from fifteenth-century panel paintings to contemporary digital illustration; twenty-six were AI-generated, made with image generators and submitted by people who make AI art. More than eleven thousand people took the test. The median respondent got 60\% right. That is better than the 50\% expected from guessing, but far from reliable. The organizer also noted that the two pictures respondents most often named as their favourite were both made by AI, and that even respondents who rated AI art lowest picked AI-made favourites \citep{alexander2024results}. The test is one of many recent demonstrations that generative models have crossed a perceptual threshold. People now fail to separate AI-made faces from photographs \citep{nightingale2022ai,miller2023ai}, AI-written poems from human poetry \citep{kobis2021artificial,porter2024ai}, AI-written self-presentations from human ones \citep{jakesch2023human}, and AI-generated images, audio, and text from real media across countries \citep{frank2024representative}.

When people cannot see who made something, they decide for themselves. Most research on how people value AI art, however, studies what happens when they are told. Telling people that a painting was made by AI lowers how much they like it and how beautiful, profound, creative, and valuable they find it \citep{bellaiche2023humans,millet2023defending,horton2023bias,chiarella2022investigating,ragot2020ai}, and AI labels similarly lower the appeal of music and the credibility of news headlines \citep{shank2023ai,longoni2022news}. These label studies establish a \emph{human premium} in evaluations: people rate a work more favourably when they believe a person made it. Outside the laboratory, though, art rarely arrives with a trustworthy label. People scroll past images on social media, receive poems in messages, and browse marketplaces where creators may or may not disclose their tools. There, the belief about who made a work is one that people form themselves. How does that belief relate to how much they like the work?

Two accounts predict that liking a work and believing that a person made it go together. In the first, the human premium operates on people's own beliefs: a work a person believes to be human-made is enjoyed more, so the favourite tends to be a work believed to be human. In the second, which we call the \emph{beauty-is-human} inference, people use their aesthetic response as evidence of authorship: if a work moves me, a mind with something to say must be behind it. This inference joins two well-documented tendencies. People treat their feelings as information about the object that produced them \citep{schwarz1983mood,schwarz2004metacognitive,slovic2007affect}, and they believe that art requires experience and intention, capacities they grant to people more than to machines \citep{gray2007dimensions,newman2012art,castelo2019task,messingschlager2025mind}. Two further processes could also link the two judgments: perceptions of a work that raise both, such as a sense of intention, and the pressure to answer consistently.

Laboratory studies have documented the association. Images, music, and haiku that participants believed to be machine-made received lower ratings \citep{chamberlain2018putting,shank2023ai,hitsuwari2023does}, including when the order of rating and categorizing was randomized \citep{chamberlain2018putting}. Liking ratings fell as believed AI origin rose in mixed models with participant and image effects \citep{grassini2024understanding}. Porter and Machery proposed that readers take their preference for a poem as evidence that a human wrote it \citep{porter2024ai}. Each of these studies involved at most a few hundred participants and fewer than a hundred works, which leaves four questions open. How large and how pervasive is the association across works when both the work and the person are held fixed? Is it confined to correctly recognized human-made works? How does it relate to people's stated opinions of AI art, their reported detection strategies, and their familiarity with art? And does it appear with continuous ratings recorded before the source judgment in another medium?

We address these questions with two public datasets. Study~1 analyses the full response file of the AI Art Turing Test: \NJudg{} human-or-AI judgments by \NPeople{} people, who were also asked to name a favourite picture, report a main detection strategy, rate their familiarity with human and AI art, and give their opinion of AI art. Because every respondent judged the same 50 pictures, we can compare how a person judged their favourite with how other people judged that same picture, and with how the same person judged the other pictures. Study~2 reanalyses a published poetry experiment \citep{porter2024ai} in which participants rated ten poems on fourteen qualities \emph{before} judging who wrote each one. Two further arms of that experiment told participants the author, which provides a benchmark for the size of a label effect on the same ratings. Free-text explanations that people wrote for some of their judgments provide exploratory material on the reasons they give.

Among the \NFav{} respondents who judged their own favourite, a picture was \Main{} percentage points (pp) more likely to be judged human-made when it was their favourite, holding the respondent and the picture fixed. Respondents called their own favourite human \FavCalledH{} of the time, against \NonfanSame{} for the same pictures among everyone else. The association was similar for AI-made and human-made favourites, was positive for every one of the 50 pictures, and appeared under every reported detection strategy. Respondents most negative about AI art chose AI-made favourites less often than the most positive respondents (\AttFavAIOne{} against \AttFavAIFive{}). However, the AI-made pictures they chose were ones that other respondents also tended to judge human, and they judged the same pictures human more often. Roughly a third to two fifths of respondents in every opinion group (\JointMin{} to \JointMax{}) therefore named an AI-made favourite that they believed a human had made. In Study~2, a poem's rating predicted the source judgment that followed it, by \PSlope{}~pp per point on a seven-point scale, for AI-written and human-written poems alike, even though AI-written poems were rated higher. The rating gap between poems believed human and poems believed AI was of similar size to the effect of telling people who wrote them.

The study makes two empirical contributions and one conceptual one. First, it characterizes the association for each work in a shared set. With thousands of people judging the same 50 pictures, the association can be estimated picture by picture: it was positive for every picture and similar for AI-made and human-made works. It remained large after adjustment for shared appearance and recognition, and when uncertainty from the choice of pictures was taken into account. Second, it decomposes how stated opinions of AI art relate to choices and beliefs. Negative opinions went with fewer AI-made favourites but with more human attribution among those favourites, both because the AI-made favourites chosen looked human to everyone and because the same pictures were judged human more often. An AI-made favourite believed to be human therefore remained common in every opinion group. The poetry reanalysis and the exploratory analysis of explanations set the scope of these findings: the same pattern appears in another medium with ratings recorded first, and some people state the beauty-is-human inference in their own words. Third, the study lays out the competing explanations of the association, identifies which of their predictions the data can test, and specifies the experiments that would separate them. The findings bear on disclosure of AI-generated content, on advice for detecting it, and on research that asks people both to rate a work and to judge its source.

The rest of the paper proceeds as follows. Section~\ref{sec:background} reviews research on detecting AI-made work, on the human premium, and on explanations of the association, compares the closest prior studies, and states the research questions. Section~\ref{sec:method} describes the two datasets and the estimation strategy. Section~\ref{sec:results} reports the results in the order of the research questions, and Section~\ref{sec:discussion} discusses what they imply, how they might be explained, and where their limits lie.

\section{Background and research questions}
\label{sec:background}

\subsection{Telling AI-made work from human work}

\citet{turing1950computing} proposed to judge machine intelligence by whether people can tell a machine's answers from a person's. For creative work, generative models now pass versions of that test across media. Image generators produce faces that people cannot tell from photographs \citep{nightingale2022ai}, and that people sometimes judge to be real more often than real faces \citep{miller2023ai}. Language models write poems that people cannot separate from human poems \citep{kobis2021artificial}, and that they judge to be human-written more often than poems by famous poets \citep{porter2024ai}. In a survey of 3{,}002 people in three countries, most participants were simply guessing when they judged whether images, audio, or text were made by people or machines \citep{frank2024representative}. Studies of paintings reach the same conclusion: viewers classify AI-made and human-made art at or close to chance \citep{gangadharbatla2022role,samo2025artificial,grassini2024understanding}.

When people cannot see the answer, they fall back on cues, and the cues they use can be wrong in predictable ways. \citet{jakesch2023human} showed that people judge AI-written self-presentations by intuitive but flawed heuristics, such as treating first-person pronouns or mentions of family as signs of a human writer, and that these heuristics make human judgment predictable enough to exploit. \citet{miller2023ai} found that the facial attributes people read as signs of a real face were the ones that made AI faces look real, and that the people who erred most were the most confident. In art, viewers associate representational styles with humans and abstract styles with machines \citep{gangadharbatla2022role}. These studies describe cues in the stimulus. We study a response in the viewer: how much the viewer likes the work.

\subsection{The human premium}

A large literature documents that people evaluate creative work less favourably when they are told that a machine made it. Labelling the same paintings as AI-made rather than human-made lowered participants' liking and their ratings of beauty, profundity, and worth \citep{bellaiche2023humans}. It reduced perceived creativity and awe, especially among people who hold that creativity is uniquely human \citep{millet2023defending}, and it lowered evaluations even when participants said the works were indistinguishable \citep{horton2023bias}. Comparable label effects appear in brain responses to paintings \citep{kirk2009modulation}, in ratings of abstract paintings at an art fair \citep{chiarella2022investigating}, in music \citep{shank2023ai}, and in the credibility of news headlines \citep{longoni2022news}. Disclosing AI authorship after an initial author-blind rating also lowers evaluations of the same image \citep{popescu2026code}. The explanations offered for this premium converge. People value the effort behind a work \citep{kruger2004effort}, its authentic connection to a maker \citep{newman2012art,jago2019algorithms}, and the experience and agency they attribute to a human mind more than to a machine \citep{gray2007dimensions,messingschlager2025mind}. More broadly, people resist algorithms for tasks they see as subjective \citep{castelo2019task}; they welcome algorithmic advice in some domains \citep{logg2019algorithm} but abandon algorithms quickly after seeing them err \citep{dietvorst2015algorithm}.

Label studies hold the work fixed and manipulate the belief, which is what makes them clean. By design, however, they do not show how people form the belief when no label is given, or how the belief they form relates to how much they like the work. The next subsection considers why the two might go together.

\subsection{Why liking and believed authorship might go together}
\label{sec:accounts}

Four processes could link liking a work and believing that a person made it. We describe each in turn, together with the patterns it predicts in data such as ours.

\paragraph{Belief-driven evaluation.} The human premium may apply to self-formed beliefs as it does to labels: a person who suspects that a work is human-made enjoys it more. The association should then be stronger among people who value human authorship more. When a single favourite is chosen after all source judgments, people may choose among the works they have already judged human.

\paragraph{Aesthetic response as evidence.} People routinely use their feelings as information. A good mood makes life in general seem more satisfying unless its true cause is pointed out \citep{schwarz1983mood}. The ease of processing a stimulus is felt as liking and read as truth or familiarity \citep{reber2004processing,schwarz2004metacognitive}. The affect a hazard evokes shapes judgments of its risks and benefits \citep{slovic2007affect}, and physical attractiveness spills over into judgments of character \citep{dion1972beautiful}. If people believe that art requires a mind that feels and intends, and that machines lack such a mind \citep{gray2007dimensions,castelo2019task,messingschlager2025mind}, then being moved by a work is evidence, in their own reasoning, that a person made it. This \emph{beauty-is-human} inference predicts the association even among people who hold no aversion to AI art. It also predicts that people who explain their judgments will sometimes cite aesthetic merit as a sign of a human maker. A distinctive prediction, which requires an experiment, is that the association should weaken when the incidental source of a person's pleasure is made salient, as misattributed moods lose their influence once their cause is pointed out \citep{schwarz1983mood}. The prediction holds only if the reminder does not itself change source beliefs, attention, or how people read the work.

\paragraph{Shared perceptions of the work.} Something in a person's response to a particular work may raise both liking and the human judgment, such as recognizing it, finding it comprehensible, or sensing intention in it. Recognition of known human works predicts an association concentrated among people who had seen the works before and confined to human-made works. A perception that can arise for either source, such as a sense of intention, predicts an association for AI-made works as well.

\paragraph{Response consistency.} Having given one answer, people may give a compatible second answer. This predicts the association whichever judgment is reported first.

Observational data cannot separate these processes completely, and they are not mutually exclusive. They can, however, be checked against the patterns that each predicts, which is what our design is built to do.

\subsection{Prior evidence on the association}

Table~\ref{tab:prior} compares the closest prior studies with ours. \citet{chamberlain2018putting} randomized whether 65 participants first categorized 60 images as computer-generated or man-made or first rated how attractive they were. Images that participants categorized as computer-generated were rated lower, and the order of the two tasks made no difference. In \citet{shank2023ai}, listeners first judged whether human-composed music excerpts were composed by an AI or a human and then rated their liking; liking predicted the judgment in mixed models with participant and excerpt effects. Evaluators of haiku gave lower scores to poems they felt were AI-generated \citep{hitsuwari2023does}. \citet{grassini2024understanding} asked 201 participants to rate 40 images on liking and on a continuous scale of likely origin; images believed to be AI-made were liked less, with participant and image effects in the models, and personality and creative identity moderated the pattern. \citet{porter2024ai} found that ratings of poem qualities accounted for participants' tendency to judge AI-written poems as human, and proposed that readers interpret their preference as evidence of human authorship. Our Study~2 reanalyses the condition of their Study~2 in which participants were not told who wrote the poems (the told-nothing condition).

\begin{table}[t]
\centering
\caption{The closest prior studies of liking and believed authorship, and this study.}
\label{tab:prior}
\scriptsize
\setlength{\tabcolsep}{3pt}
\begin{tabular}{>{\raggedright\arraybackslash}p{0.13\linewidth}>{\raggedright\arraybackslash}p{0.155\linewidth}>{\raggedright\arraybackslash}p{0.1\linewidth}>{\raggedright\arraybackslash}p{0.085\linewidth}>{\raggedright\arraybackslash}p{0.13\linewidth}>{\raggedright\arraybackslash}p{0.14\linewidth}>{\raggedright\arraybackslash}p{0.12\linewidth}}
\toprule
Study & Medium; participants; works & Liking measure & Source judgment & Order & Adjustment for people and works & Moderators \\
\midrule
\citet{chamberlain2018putting}, Study 1 & Images; 65; 60 (30 computer-generated) & Attractiveness, 1--7 & Binary & Randomized between participants (blocks) & Mean ratings by own categorization, within participants & Order; art education \\
\citet{shank2023ai}, Study 1 & Music; 146 and 149; 20 human-composed excerpts per genre & Liking, 7 points & Binary & Source judgment, then liking & Random intercepts for participants and excerpts & Genre \\
\citet{grassini2024understanding} & Images; 201; 40 (20 AI-made) & Liking, 0--100 & Continuous, 0--100 & Liking listed before origin & Random intercepts for participants and images & Personality, creative identity, technology attitudes \\
\citet{porter2024ai}, Study 2 & Poems; 230 of 696 guessed the source; 10 (5 AI-generated) & 14 qualities, 7 points & Binary & Ratings, then source judgment & Random intercepts for participants and poems (factor scores) & Framing arms \\
This paper, Study 1 & Images; \NFav{} of \NPeople{}; 50 (\NAIImg{} AI-made) & One favourite & Binary & Favourite after all judgments & Respondent and picture fixed effects; picture-specific estimates & Opinion of AI art, strategy, familiarity \\
This paper, Study 2 & Reanalysis of \citet{porter2024ai}, Study 2 & Quality, 1--7 & Binary & Ratings, then source judgment & Participant and poem fixed effects & Source of poem; label benchmark \\
\bottomrule
\end{tabular}
\end{table}

These studies establish that liking and believed human authorship go together across media, and Chamberlain et al.'s order manipulation shows that the link does not depend on which judgment comes first. What they could not establish, with fewer than a hundred works and at most a few hundred participants, is how the association varies across works, whether it is confined to human-made works, and how it relates to stated opinions of AI art and reported detection strategies. A large sample of people judging the same set of works can answer these questions with person and work fixed effects. These are a stricter adjustment than the random intercepts of earlier models, because they do not assume that person or work effects are unrelated to favourite status.

\subsection{Opinions, strategies, and familiarity}

An opinion of AI art is a judgment about a category. Someone who dislikes AI art in general can still love an exceptional AI-made picture, so choosing an AI-made favourite does not contradict a negative opinion. What the opinion predicts is an empirical question with two parts: whether people who dislike AI art choose fewer AI-made favourites, and whether they are more likely to believe that the favourite they chose was made by a person. Stated attitudes are often weak guides to behaviour toward the object of the attitude \citep{lapiere1934attitudes,wicker1969attitudes,ajzen1977attitude}, and opinions may relate to choices and beliefs in different ways. If belief-driven evaluation drives the association, it should be strongest among people with the most negative opinion of AI art.

Detection strategies and familiarity with art are natural places to look for differences between people. Those who report looking for technical errors may rely less on how much they like a picture. Those who know art well have seen more human work, and those who make AI art know first-hand what generators can produce. Either kind of experience could weaken the association.

\subsection{Research questions}

The accounts and gaps reviewed above lead to five research questions. The first two concern the association itself, the third concerns differences between people, the fourth concerns a second medium, and the fifth is exploratory.

\begin{description}
\item[RQ1.] Holding the work and the person fixed, how strongly is liking a work associated with judging it human-made, and how consistent is the association across works?
\item[RQ2.] Is the association confined to human-made works, or does it also appear for AI-made works?
\item[RQ3.] How do stated opinions of AI art relate to the source of people's favourites, to their beliefs about those favourites, and to the strength of the association? How does the association vary with reported detection strategies and familiarity with art?
\item[RQ4.] Does the association appear in another medium when liking is rated before each source judgment, and how does its size compare with the effect of an explicit authorship label?
\item[RQ5.] What reasons do people give when they explain their source judgments? (exploratory)
\end{description}

\section{Data and design}
\label{sec:method}

We answer the research questions with two public, de-identified datasets, summarized in Table~\ref{tab:data}. Study~1 provides the main evidence; Study~2 reverses the order in which liking and the source judgment are measured.

\begin{table}[t]
\centering
\caption{Data sources.}
\label{tab:data}
\small
\begin{tabular}{p{0.2\linewidth}p{0.36\linewidth}p{0.36\linewidth}}
\toprule
 & Study 1: AI Art Turing Test & Study 2: poetry experiment \\
\midrule
Source & Public response file of an online test run by a blog \citep{alexander2024results,alexander2024data} & Open data of \citet{porter2024ai}, Study 2 \citep{porter2024data} \\
Works & 50 pictures: \NHumanImg{} human-made, \NAIImg{} AI-made & 10 poems: 5 human-written, 5 AI-generated \\
Respondents & \NPeople{}; analysis sample of \NFav{} who named and judged a favourite & \PNAll{}; analysis sample of \PNPeople{} in the condition that judged the source \\
Judgments & \NJudg{} source judgments (\NRows{} in the analysis sample) & \PNJudg{} source judgments \\
Liking & One favourite picture, named after all judgments & Fourteen 7-point ratings per poem, given before each judgment \\
Explanations & \TNImg{} free-text explanations for six pictures & \TNPoem{} free-text explanations from \citeauthor{porter2024ai}'s Study 1 \\
\bottomrule
\end{tabular}
\end{table}

\subsection{Study 1: the AI Art Turing Test}

The test was posted on the Astral Codex Ten blog in the autumn of 2024, and the organizer later published the de-identified responses with a summary of results \citep{alexander2024results,alexander2024data}. It showed 50 pictures in four broad styles: Renaissance and older painting, nineteenth-century painting, abstract and modern art, and digital art. The human-made pictures were well-regarded works. The AI-made pictures came from people who make AI art; the organizer screened out images with obvious defects such as garbled text or malformed hands. Each picture was identified by a short descriptive name, such as ``Riverside Cafe'' or ``Wounded Christ''. For each picture, respondents answered whether it was made by a human or by AI. For six pictures they also rated their confidence (recorded values range from 50 to 100) and could explain their answer in a sentence. One human-made portrait was not shown to every respondent (the organizer reports an error in adding it to the form) and was judged by 8{,}556 of them.

After the pictures, respondents named three pictures from a list: the one they were most confident was human-made, the one they were most confident was AI-made, and their favourite (``Which picture is your favorite?''). They then reported their familiarity with human art and with AI art (four levels each, from ``basically zero'' to ``[AI] art is either my job or my greatest passion, I think I qualify as an expert in it''), how many pictures they had seen before, the share they thought they had right, the main strategy they had used (six options, listed in Table~\ref{tab:groups}), and their opinion of AI art ``on an artistic level'' on a scale from 1 (most negative) to 5 (most positive). The file includes a scored copy of each response, from which we recovered the true source of every picture; the recovered source agreed with the scoring for every one of the \NJudg{} judgments.

Our analysis sample consists of the \NFav{} respondents who named a favourite and also judged that picture, and their \NRows{} judgments. Of the \FlowExcl{} excluded respondents (\FlowExclPct{}), \FlowNoFav{} did not name a favourite and \FlowNotJudged{} did not judge the picture they named. Excluded respondents had answered fewer pictures (\FlowAnsEx{} against \FlowAnsIn{} on average), were slightly less accurate (\FlowAccEx{} against \FlowAccIn{}), and more often left the closing questions unanswered: \FlowOpMissEx{} of them gave no opinion of AI art, against \FlowOpMissIn{} of included respondents (Appendix Table~\ref{tab:flow}). The main outcome is whether a respondent judged a picture to be human-made. The main predictor is whether the picture was the respondent's favourite.

\subsection{Study 2: ratings recorded before the source judgment}

In Study~1, the favourite was named after all source judgments. Study~2 reverses the order. We reanalyse Study~2 of \citet{porter2024ai}, whose data and code are public \citep{porter2024data}. Participants read ten poems: five by human poets (Chaucer, Shakespeare, Dickinson, Eliot, and Lasky) and five generated by ChatGPT~3.5 in the styles of five other poets. They rated each poem on fourteen qualities on 7-point scales, including how good, beautiful, moving, and meaningful it was. The consent page described the study as being about perceptions of authorship in poetry. Participants were randomly assigned to one of three conditions. In one, each poem was introduced as ``written by'' a named poet; in another, as ``written by an AI program''; in the third, no author was given, and after rating each poem participants answered ``Do you think this poem was written by a human, or generated by an AI program?'' Our main sample is the \PNPeople{} participants of the told-nothing condition and their \PNJudg{} judgments. Our liking measure is the rating of how good the poem was. The fourteen ratings are strongly related measures of evaluation, and we report the others as supporting detail rather than as separate tests. The \PNLabel{} participants in the two labelled conditions provide a benchmark for the effect of an authorship label on the same ratings. That label combines the claim of human authorship with the name of a known poet.

\subsection{Explanations}

For Study~1, we use the \TNImg{} explanations that \TNImgPeople{} respondents wrote for the six pictures with an explanation prompt. For poetry, we use the \TNPoem{} optional explanations that \TNPoemPeople{} participants in Study~1 of \citet{porter2024ai} wrote for their source judgments. That study asked for judgments and confidence but not ratings, so its explanations describe reasons people give in general rather than the reasons behind the ratings of Study~2. We coded each explanation with a transparent keyword list for six kinds of reasons. Three point toward what a work expresses: aesthetic or emotional \emph{merit} (beauty, soul, emotion, meaning, creativity, authenticity), \emph{intention} (intentional, deliberate, thoughtful), and personal \emph{liking} (``I like it'', ``favourite''). Three point toward how it was made or what it lacks: \emph{technical cues} (hands, errors, inconsistencies, things that make no sense), \emph{polish} (``too perfect'', ``too clean'', ``too smooth''), and \emph{lack of merit} (generic, soulless, boring, clich\'ed). A merit word preceded within three words by a negation such as ``no'', ``lacks'', or ``without'' counts as lack of merit; other, non-negated merit words in the same explanation still count as merit. Categories are not exclusive. The full lists and randomly drawn examples are in Appendix~\ref{app:text}. This analysis is exploratory and descriptive, and it plays no part in the tests of RQ1--RQ4.

\subsection{Estimation}

For Study~1, we estimate linear probability models of the form
\begin{equation}
\text{JudgedHuman}_{ij} = \beta\,\text{Favourite}_{ij} + \alpha_i + \gamma_j + \varepsilon_{ij},
\label{eq:main}
\end{equation}
where $i$ indexes respondents and $j$ pictures. The respondent fixed effect $\alpha_i$ absorbs how often a person calls pictures human in general, and the picture fixed effect $\gamma_j$ absorbs how human a picture looks to everyone. The coefficient $\beta$ is therefore the difference, in percentage points, between how likely a person is to call a picture human when it is their favourite and what their general tendency and the picture's general appearance predict. It is an association: the fixed effects remove stable differences between respondents and between pictures, not a particular respondent's particular response to a particular picture. We absorb the fixed effects exactly and cluster standard errors by respondent. A logit model with picture indicators and the respondent's leave-one-out share of ``human'' judgments, a stand-in for respondent fixed effects, gives the same answer (Table~\ref{tab:specs}).

Clustering by respondent treats the 50 pictures as fixed. To express uncertainty that includes variation across pictures, we also report standard errors clustered by respondent and picture, estimates that leave out one picture at a time (dropping the picture and the respondents whose favourite it was), and picture-specific estimates summarized with a random-effects model (DerSimonian--Laird). These describe the 50 pictures in the test; they are not predictions for pictures selected in other ways.

To ask whether the association depends on the true source, we interact the favourite indicator with whether the picture was AI-made. Differences between groups of respondents come from a pooled model with respondent fixed effects and group-by-picture fixed effects. That model reproduces the separate within-group estimates reported in the tables and gives their difference with the joint covariance.

A favourite can look human to its fans because of its style or other features shared with other pictures, rather than because they like it. People who like classical painting may call every classical painting human. We address this with \emph{co-judgment clusters}: groups of pictures that the same people tend to call human, after removing each person's overall tendency. The clusters capture style and any other shared appearance, and they are not an independent measure of artistic style. Replacing $\alpha_i$ with a fixed effect for each respondent--cluster pair compares a favourite only with the respondent's judgments of pictures in the same cluster. Because the clusters are built from the same judgments, we also estimate them on a random half of respondents and apply them to the other half. Finally, we estimate how much the favourite association spills over to the other pictures in the favourite's cluster.

For RQ2, we also summarize the judgments with signal detection measures \citep{stanislaw1999calculation}. Treating ``human-made'' as the signal, the hit rate $H$ is the share of human-made pictures judged human and the false-alarm rate $F$ is the share of AI-made pictures judged human. Sensitivity is $d' = z(H) - z(F)$, and the criterion $c = -[z(H) + z(F)]/2$ measures the tendency to answer ``human'' (negative values mean a lenient criterion). We compute both for favourites and for the same pictures judged by respondents who did not choose them, weighting pictures by their number of fans, with 95\% intervals from 1{,}000 bootstrap resamples of respondents. Because each respondent has only one favourite, these are aggregate descriptions across respondents, not estimates of a within-person change.

For Study~2, we estimate Equation~\eqref{eq:main} with participant and poem fixed effects. Liking enters either as the rating itself, our primary poetry measure, or as an indicator for the poem or poems a participant rated highest (ties included). To compare the size of the association with that of a label, we also estimate the reverse model: the rating of a poem regressed on whether the participant judged it human, with the same fixed effects. We compare that rating gap with the difference in ratings between participants told that the poems were written by named poets and participants told they were written by an AI program, with poem fixed effects and standard errors clustered by participant. The two quantities are different estimands. The first is an association with a person's own judgment, and the second is the effect of assigned information in different participants. We compare them only for scale.

The primary analyses are the favourite association of Equation~\eqref{eq:main} in Study~1 and the rating association in Study~2. The subgroup, attitude, and explanation analyses are secondary and exploratory. None of the analyses was preregistered, and inference refers to the works in each dataset.

\section{Results}
\label{sec:results}

We report the results in the order of the research questions. As a starting point, respondents in Study~1 judged \AccMean{} of the pictures correctly on average (median \AccMedian{}). AI-made pictures were judged human \HumanRateAI{} of the time and human-made pictures \HumanRateH{} of the time, so most respondents could separate the two sources only weakly.

\subsection{Favourites are judged human more often (RQ1)}

\begin{figure}[t]
\centering
\includegraphics[width=\linewidth]{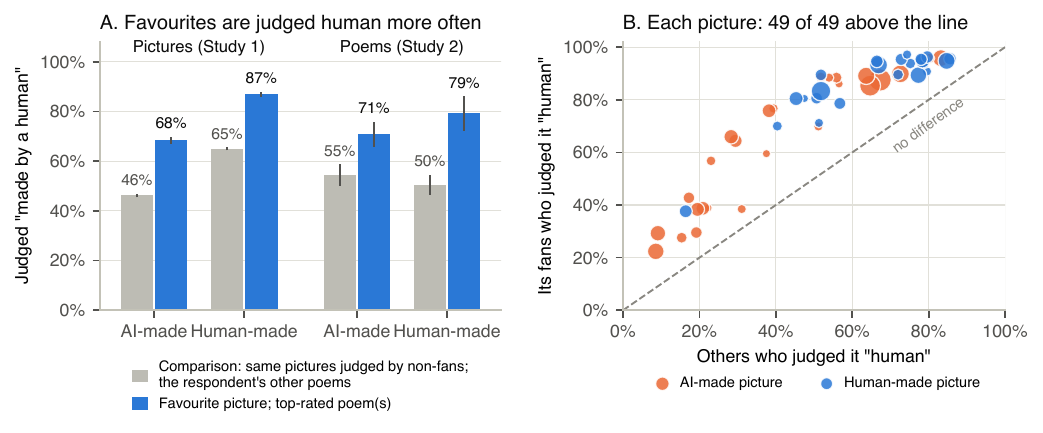}
\caption{Favourites are judged human more often. (A) Share of judgments ``made by a human'' for each respondent's favourite picture and for the same pictures judged by respondents who did not choose them (Study~1, pictures weighted by their number of fans), and for each respondent's top-rated poem or poems and their other poems (Study~2), by true source. Bars show 95\% intervals from 1{,}000 bootstrap resamples of respondents (participants for poems). These are raw shares; Table~\ref{tab:specs} and Table~\ref{tab:poetry} give the fixed-effects estimates. (B) Each dot is one of the \ImgNGe{} pictures with at least 30 fans: the share of its fans (vertical) and of all other respondents (horizontal) who judged it human. Dot size is proportional to the number of fans; the dashed line marks equal shares.}
\label{fig:coupling}
\end{figure}

Respondents judged their own favourite to be human-made \FavCalledH{} of the time. The same pictures were judged human-made \NonfanSame{} of the time by respondents who did not choose them (Figure~\ref{fig:coupling}A). Holding the respondent and the picture fixed, Equation~\eqref{eq:main} gives a favourite association of \Main{}~pp (SE \MainSE{}; 95\% CI \MainLo{} to \MainHi{}). Clustering standard errors by picture as well as by respondent widens the interval to \MainLoTwo{} to \MainHiTwo{}~pp. The logit model gives an odds ratio of \LogitOR{} and an average change of \LogitAME{}~pp in the probability of a ``human'' judgment between non-favourite and favourite (Table~\ref{tab:specs}).

The association holds for each picture, not only on average. Every one of the \ImgNGe{} pictures with at least 30 fans was judged human more often by its fans than by everyone else (Figure~\ref{fig:coupling}B). Picture-specific estimates with respondent and picture fixed effects were positive for all 50 pictures, ranging from \PicMin{} to \PicMax{}~pp; \PicHolm{} remained significant after Holm correction for 50 tests (Appendix Table~\ref{tab:pictures}). A random-effects summary of these estimates gives a mean of \PicMu{}~pp (SE \PicMuSE{}) with an estimated between-picture standard deviation of \PicTau{}~pp. Leaving out one picture at a time, together with the respondents whose favourite it was, moves the estimate only between \LooMin{} and \LooMax{}~pp. Even pictures that most respondents considered obviously AI-made were called human more often by their fans (the AI-made \emph{Ancient Gate}: \ExAGFans{} of fans against \ExAGOthers{} of others), and pictures that most respondents considered human, whether human-made or not, were called human by their fans almost unanimously (the AI-made \emph{Paris Scene}: \ExPSFans{} against \ExPSOthers{}).

\begin{table}[t]
\centering
\caption{The favourite association in Study~1 across specifications. Linear probability models of judging a picture human-made, with respondent and picture fixed effects unless stated otherwise; standard errors clustered by respondent unless stated otherwise. Co-judgment clusters group pictures that the same people tend to call human. $^a$The logit model uses picture indicators and the respondent's leave-one-out share of ``human'' judgments in place of respondent fixed effects. $^b$Each estimate drops one picture and the respondents whose favourite it was, so the number of respondents varies.}
\label{tab:specs}
\small
\begin{tabular}{lrr}
\toprule
Specification & Estimate, pp (SE) & Respondents \\
\midrule
Baseline: respondent and picture fixed effects & $+$22.1 (0.4) & 9{,}152 \\
\quad Same, standard error clustered by respondent and picture & $+$22.1 (1.3) & 9{,}152 \\
\quad Favourite was AI-made & $+$21.8 (0.6) & 5{,}092 \\
\quad Favourite was human-made & $+$22.4 (0.5) & 4{,}060 \\
Respondent $\times$ co-judgment-cluster fixed effects (4 clusters) & $+$20.0 (0.4) & 9{,}152 \\
Respondent $\times$ co-judgment-cluster fixed effects (6 clusters) & $+$19.7 (0.4) & 9{,}152 \\
Respondent $\times$ co-judgment-cluster fixed effects (8 clusters) & $+$19.0 (0.4) & 9{,}152 \\
\quad Clusters estimated on the other half of respondents (4 clusters) & $+$19.9 (0.6) & 4{,}496 \\
\quad Spillover to other pictures in the favourite's cluster (4 clusters) & $+$2.9 (0.2) & 9{,}152 \\
Had seen none of the pictures before & $+$22.6 (0.6) & 4{,}704 \\
Favourite differs from most-confident-human pick & $+$20.7 (0.4) & 8{,}077 \\
Judged 49 or 50 pictures & $+$22.2 (0.4) & 8{,}727 \\
Overall human share between 40\% and 60\% & $+$22.5 (0.4) & 7{,}122 \\
Excluding the late-added picture & $+$21.9 (0.4) & -- \\
Excluding the six pictures with confidence prompts & $+$21.7 (0.4) & -- \\
\midrule
Leave one picture out$^b$: range of estimates & 21.5 to 22.5 & 8{,}510--9{,}129 \\
Logit$^a$, average change from non-favourite to favourite (odds ratio 3.87) & $+$23.6 (0.4) & 9{,}152 \\
\bottomrule
\end{tabular}

\end{table}

Three alternative explanations predict that the association should shrink under particular comparisons, and it shrinks little. The first is shared appearance: people who like a style may call every picture in that style human. Comparing each favourite only with the respondent's judgments of pictures in the same co-judgment cluster leaves an association of \StyleFour{}~pp with four clusters and \StyleEight{}~pp with eight. With clusters estimated on one half of the respondents and applied to the other half, the estimate is \HeldFour{}~pp (\HeldBase{}~pp without clusters in that half). Fans did call the other pictures in their favourite's cluster human slightly more often, by \SpillFour{}~pp, but the association with the favourite itself was about \SpillRatio{} times as large. The second is recognition: a person who has seen a famous painting before may both love it and know it is human. Among the \RobSeenN{} respondents who had seen none of the pictures before, the association was \RobSeen{}~pp. The third is that the favourite might simply duplicate the picture the person was most confident was human. Excluding respondents whose favourite and most-confident-human picture coincided leaves \RobNoMch{}~pp. The estimate is also unchanged among respondents who judged all pictures, among those whose overall rate of ``human'' answers was between 40\% and 60\%, and without the late-added picture or the six pictures with confidence prompts (Table~\ref{tab:specs}).

Favourite status was also associated with confidence. For the six pictures with a confidence rating, fans were \ConfUns{} points more confident in their answer than other respondents judging the same pictures, holding respondent and picture fixed (SE \ConfUnsSE{}). The difference appeared when fans judged their favourite human (\ConfUnsH{} points, SE \ConfUnsHSE{}) and not when they judged it AI-made (\ConfUnsAI{}, SE \ConfUnsAISE{}); the difference between the two is \ConfContrast{} points (95\% CI \ConfContrastLo{} to \ConfContrastHi{}).

\subsection{The association appears for AI-made and human-made works alike (RQ2)}

The association could arise because favourites tend to be human works that people correctly recognize. In that case it would be confined to human-made favourites, and it was not. The favourite association was \MainAI{}~pp (SE \MainAISE{}) when the favourite was AI-made and \MainH{}~pp (SE \MainHSE{}) when it was human-made. The difference is \SrcContrast{}~pp, with a 95\% CI of \SrcContrastLo{} to \SrcContrastHi{} when standard errors are clustered by respondent and \SrcContrastLoTwo{} to \SrcContrastHiTwo{} when they are clustered by respondent and picture. The random-effects means of the picture-specific estimates were \PicMuAI{}~pp for AI-made and \PicMuH{}~pp for human-made pictures. Respondents were more accurate on human-made favourites than on other human-made pictures, by \CorrH{}~pp, and less accurate on AI-made favourites than on other AI-made pictures, by \CorrAIAbs{}~pp. The association is thus not confined to correctly recognized human works. It also occurs for AI-made works, consistent with evaluative or perceptual cues that can arise for either source.

Signal detection measures point the same way: favourites drew more ``human'' answers whatever their source. For favourites, the hit rate was \FavCalledHH{} and the false-alarm rate \FavCalledHAI{}; for the same pictures judged by other respondents, the rates were \NonfanSameH{} and \NonfanSameAI{}. The criterion was more lenient toward ``human'' for favourites ($\Delta c =$ \DeltaC{}, 95\% CI \DeltaCLo{} to \DeltaCHi{}). Sensitivity was somewhat higher ($\Delta d' =$ \DeltaD{}, 95\% CI \DeltaDLo{} to \DeltaDHi{}), while the difference between hit and false-alarm rates was unchanged (\DeltaHmF{}~pp, 95\% CI \DeltaHmFLo{} to \DeltaHmFHi{}). The two discrimination measures differ because $d'$ transforms rates nonlinearly.

Favourite status was also not positively diagnostic of human authorship in this set of pictures. AI-made pictures made up \FavAIShare{} of respondents' favourites, against \FavAIOthers{} of the other pictures the same respondents judged (difference \FavAIPaired{}~pp, 95\% CI \FavAIPairedLo{} to \FavAIPairedHi{}), and \TopTenAI{} of the 10 most-chosen favourites were AI-made. The most-chosen favourite, the AI-made \emph{\TopImg{}} (\TopImgFavN{} fans), was judged human by \TopImgHumanRate{} of all respondents. The favourite also coincided with the respondent's most-confident-human picture \FavEqMch{} of the time, \FavEqMchRatio{} times the \FavEqMchChance{} expected if the two choices were unrelated. It coincided with the most-confident-AI picture \FavEqMca{} of the time, below the \FavEqMcaChance{} expected. Of the pictures that respondents were most confident a human had made, \MchShareAI{} were AI-made; the AI-made \emph{\MchTopAIImg{}} was the \MchTopAIRank{} common choice, named by \MchTopAIN{} respondents.

\subsection{Opinions of AI art, choices, and beliefs (RQ3)}

\begin{figure}[t]
\centering
\includegraphics[width=\linewidth]{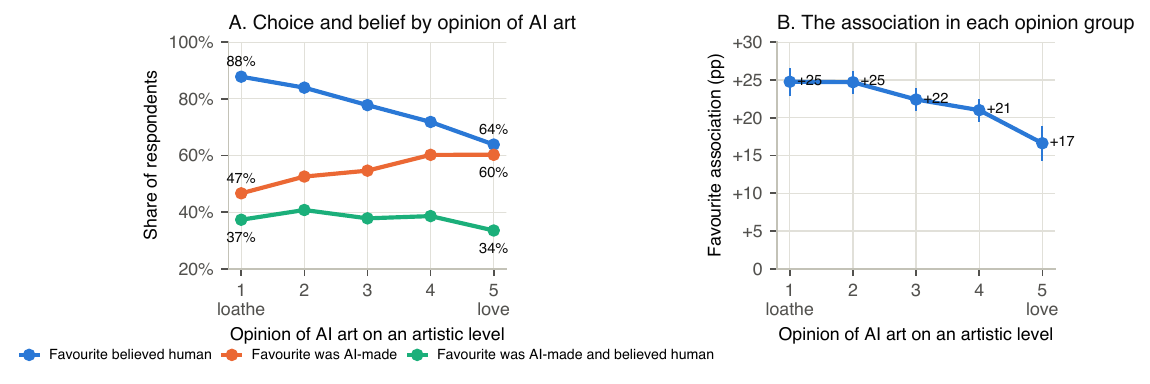}
\caption{Choice and belief by opinion of AI art (Study~1). (A) By opinion of AI art, the share of respondents whose favourite was AI-made, whose favourite they believed human-made, and whose favourite was AI-made and believed human-made. (B) The favourite association within each opinion group, with 95\% confidence intervals.}
\label{fig:attitudes}
\end{figure}

We next ask how respondents' opinions of AI art relate to the pictures they chose and to what they believed about them. Opinions were spread across the scale: of the \NOpinion{} respondents who answered, \OpNeg{} were negative (1 or 2), \OpNeu{} neutral, and \OpPos{} positive (4 or 5). Table~\ref{tab:attitudes} and Figure~\ref{fig:attitudes} relate these opinions to whether the respondent's favourite was AI-made (what they chose), whether they judged it human-made (what they believed), and the favourite association.

\begin{table}[t]
\centering
\caption{Opinion of AI art, choice of favourite, and belief about the favourite (Study~1). Respondents who named and judged a favourite. The row on other respondents gives the share of all other respondents who judged the chosen AI-made favourites human. Accuracy is computed for the same respondents. The favourite association is estimated within each opinion group as in Equation~\eqref{eq:main}.}
\label{tab:attitudes}
\small
\begin{tabular}{lrrrrr}
\toprule
 & \multicolumn{5}{c}{Opinion of AI art on an artistic level} \\
\cmidrule(lr){2-6}
 & 1 (loathe) & 2 & 3 & 4 & 5 \\
\midrule
Respondents & 1{,}106 & 1{,}891 & 2{,}181 & 2{,}620 & 1{,}270 \\
Favourite was AI-made & 47\% & 53\% & 55\% & 60\% & 60\% \\
Favourite believed human-made & 88\% & 84\% & 78\% & 72\% & 64\% \\
AI-made favourite believed human-made & 80\% & 78\% & 69\% & 64\% & 56\% \\
\quad Same pictures judged human by others & 54\% & 52\% & 47\% & 43\% & 38\% \\
Favourite AI-made \emph{and} believed human & 37\% & 41\% & 38\% & 39\% & 34\% \\
AI-made among favourites believed human & 43\% & 49\% & 49\% & 54\% & 53\% \\
Accuracy, all judged pictures & 63.9\% & 61.6\% & 60.5\% & 59.8\% & 60.3\% \\
Accuracy on an AI-made favourite & 20\% & 22\% & 31\% & 36\% & 44\% \\
Favourite association, pp (SE) & $+$24.8 (1.0) & $+$24.7 (0.8) & $+$22.4 (0.8) & $+$21.0 (0.8) & $+$16.6 (1.2) \\
\bottomrule
\end{tabular}

\end{table}

Opinion went with choice. Among respondents most negative about AI art, \AttFavAIOne{} picked an AI-made favourite; among the most positive, \AttFavAIFive{} did, an increase of \SlopeFavAI{}~pp per step on the five-point scale (SE \SlopeFavAISE{}). Opinion also went with belief. The most negative respondents judged their favourite human-made \AttCalledHOne{} of the time, and the most positive \AttCalledHFive{} of the time. When the favourite was in fact AI-made, \AttAIBelHOne{} of the most negative respondents believed a human had made it, against \AttAIBelHFive{} of the most positive, a gap of \RawPicGap{}~pp.

Two patterns account for that gap. First, the groups chose different AI-made pictures. The AI-made favourites of the most negative respondents were judged human by \AttAIOthersHOne{} of all other respondents, against \AttAIOthersHFive{} for those of the most positive. People who disliked AI art, when they picked an AI-made favourite, tended to pick one that looked human to everyone.

Second, the groups judged the same pictures differently. Holding the favourite picture fixed, the most negative respondents were \SamePicGap{}~pp (95\% CI \SamePicGapLo{} to \SamePicGapHi{}) more likely than the most positive to believe their AI-made favourite was human-made. The groups overlapped substantially: \OverlapPics{} AI-made pictures that both groups chose at least ten times accounted for \OverlapShareOne{} and \OverlapShareFive{} of their AI-made favourites. Restricted to these shared pictures, the same-picture gap is \SharedGap{}~pp (95\% CI \SharedGapLo{} to \SharedGapHi{}; \SharedGapN{} respondents). Across the whole scale, holding the picture fixed reduces the decline in believed human authorship of AI-made favourites from \AttRawPicSlope{} to \AttSamePicSlope{}~pp per step (SE \AttSamePicSlopeSE{}). These comparisons describe the gap; they do not divide it into causal components, because opinions were reported after the task and choosing a favourite is itself part of the observed behaviour.

The share of respondents whose favourite was AI-made \emph{and} believed human-made is the product of the two shares: the probability of an AI-made favourite times the probability of judging an AI-made favourite human. The first rose with opinion and the second fell, so the product stayed between \JointMin{} and \JointMax{} across the five groups (\AttAIAndHOne{} among the most negative, \AttAIAndHFive{} among the most positive, and \AIAndHAll{} overall). More than a third of the people who held the lowest opinion of AI art named an AI-made picture as their favourite and believed a person had made it. Of the favourites these respondents believed to be human-made, \AttBelHIsAIOne{} were AI-made.

Respondents who disliked AI art were not worse at the task. The most negative respondents were the most accurate group overall (\AttAccOne{}, against \AttAccFive{} for the most positive), with higher sensitivity ($d' =$ \AttDOne{} against \AttDFive{}) and a similar overall criterion (\AttCOne{} against \AttCFive{}). When their favourite was AI-made, they identified it as AI-made \AttAccAIFavOne{} of the time, against \AttAccAIFavFive{} for the most positive respondents.

The favourite association appeared in every opinion group (Figure~\ref{fig:attitudes}B). It was \AttTwfeOne{}~pp among the most negative respondents and declined by \AttSlope{}~pp per step (95\% CI \AttSlopeLo{} to \AttSlopeHi{}) to \AttTwfeFive{}~pp (SE \AttTwfeSEFive{}) among the most positive; the difference between the extreme groups is \AttFiveContrast{}~pp (95\% CI \AttFiveContrastLo{} to \AttFiveContrastHi{}). A stronger association among people who dislike AI art is what belief-driven evaluation predicts. The association that remained among people who rate AI art most highly, about two thirds as large, is what an account that does not depend on aversion predicts.

\subsection{Reported strategies and familiarity (RQ3)}

\begin{figure}[t]
\centering
\includegraphics[width=0.86\linewidth]{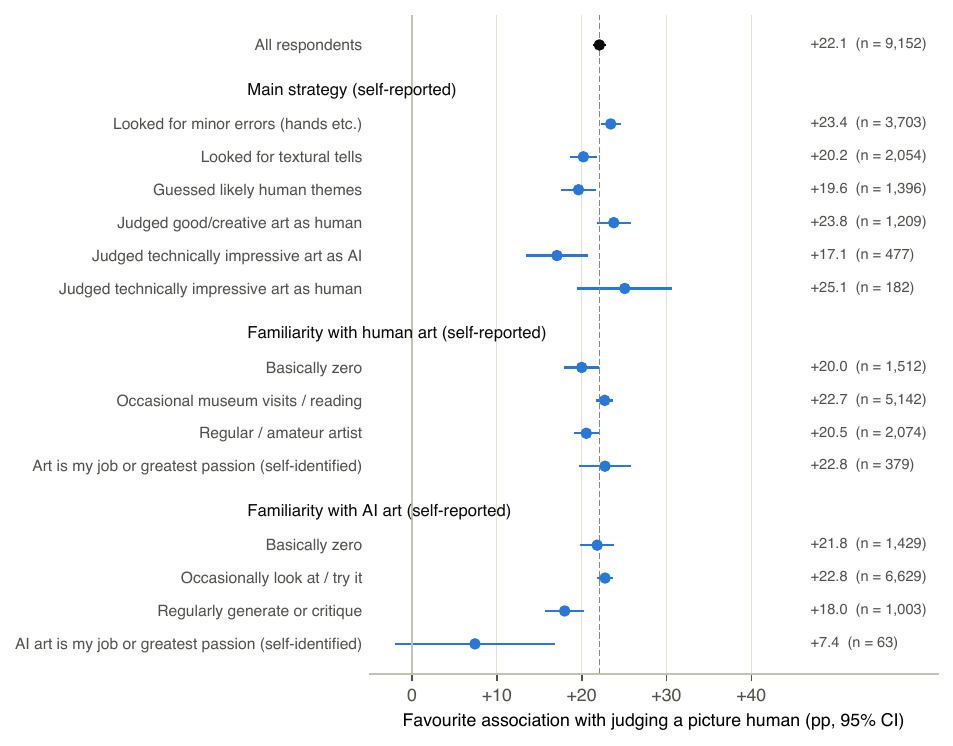}
\caption{The favourite association by reported strategy and familiarity (Study~1), with 95\% confidence intervals. The dashed line marks the association among all respondents.}
\label{fig:groups}
\end{figure}

Respondents also reported how they made their judgments and how familiar they were with art. The favourite association was large under every main strategy that respondents reported, from \StratMin{} to \StratMax{}~pp (Table~\ref{tab:groups} and Figure~\ref{fig:groups}). The most common strategy, looking for minor errors such as hands with the wrong number of fingers, went with an association of \StratErrors{}~pp (SE \StratErrorsSE{}; \StratErrorsN{} respondents). That is as large as among respondents who said they judged good or creative art as human, \StratGoodH{}~pp (SE \StratGoodHSE{}; \StratGoodHN{} respondents): the difference is \ErrGoodContrast{}~pp (95\% CI \ErrGoodLo{} to \ErrGoodHi{}). Strategies differed overall (joint test \StratWaldP{}), mainly because the association was smallest, \StratImprAI{}~pp, among respondents who said they treated technical impressiveness as a sign of AI. The association was present in every strategy group.

\begin{table}[t]
\centering
\caption{The favourite association by reported strategy and familiarity (Study~1). Accuracy is the mean share of judged pictures answered correctly. The favourite association is estimated within each group as in Equation~\eqref{eq:main}; differences between groups reported in the text come from a pooled model with group-specific picture effects.}
\label{tab:groups}
\small
\begin{tabular}{lrrr}
\toprule
 & Respondents & Accuracy & Favourite association, pp (SE) \\
\midrule
\multicolumn{4}{l}{\emph{Main strategy (self-reported)}} \\
\quad Looked for minor errors, e.g.\ hands & 3{,}703 & 61.9\% & $+$23.4 (0.6) \\
\quad Looked for textural tells & 2{,}054 & 62.3\% & $+$20.2 (0.8) \\
\quad Guessed themes humans would paint & 1{,}396 & 58.9\% & $+$19.6 (1.1) \\
\quad Judged good/creative art as human & 1{,}209 & 59.3\% & $+$23.8 (1.0) \\
\quad Judged technically impressive art as AI & 477 & 58.5\% & $+$17.1 (1.8) \\
\quad Judged technically impressive art as human & 182 & 58.7\% & $+$25.1 (2.8) \\
\multicolumn{4}{l}{\emph{Familiarity with human art (self-reported)}} \\
\quad Basically zero & 1{,}512 & 59.0\% & $+$20.0 (1.1) \\
\quad Occasionally reads about art / visits museums & 5{,}142 & 60.1\% & $+$22.7 (0.5) \\
\quad Regularly; may be an amateur artist & 2{,}074 & 63.1\% & $+$20.5 (0.8) \\
\quad Art is my job or greatest passion (self-identified expert) & 379 & 66.9\% & $+$22.8 (1.6) \\
\multicolumn{4}{l}{\emph{Familiarity with AI art (self-reported)}} \\
\quad Basically zero & 1{,}429 & 58.1\% & $+$21.8 (1.0) \\
\quad Occasionally looks at or generates & 6{,}629 & 60.9\% & $+$22.8 (0.5) \\
\quad Regularly generates or critiques & 1{,}003 & 64.9\% & $+$18.0 (1.2) \\
\quad AI art is my job or greatest passion (self-identified expert) & 63 & 63.9\% & $+$7.4 (4.8) \\
\bottomrule
\end{tabular}

\end{table}

Familiarity with human art went with accuracy but not with a weaker association. Respondents who described art as their job or greatest passion and considered themselves experts were the most accurate familiarity group (\FamHExpertAcc{}, against \FamHZeroAcc{} for those with no familiarity). Their favourite association was \FamHExpert{}~pp (SE \FamHExpertSE{}; \FamHExpertN{} respondents), with no clear difference from all other respondents (\FamHContrast{}~pp, 95\% CI \FamHContrastLo{} to \FamHContrastHi{}). Familiarity with AI art showed a different pattern at the top. Respondents who regularly generate or critique AI images showed an association of \FamAIRegular{}~pp (SE \FamAIRegularSE{}). The \FamAIExpertN{} respondents who described AI art as their job or greatest passion showed \FamAIExpert{}~pp (SE \FamAIExpertSE{}), \FamAIContrast{}~pp smaller than all other respondents (95\% CI \FamAIContrastLo{} to \FamAIContrastHi{}; $p = $ \FamAIContrastP{}).

\subsection{Poems rated before the source judgment (RQ4)}

Study~2 asks whether the same pattern appears in another medium when the order of measurement is reversed: participants rated each poem before judging who wrote it. Participants in the told-nothing condition judged the source of \PAcc{} of the poems correctly, below chance, because they judged AI-written poems human more often (\PHumanRateAI{}) than human-written ones (\PHumanRateH{}). They also rated the AI-written poems higher: \PGoodAI{} against \PGoodH{} on the seven-point question of how good the poem was. Among these ten poems, a higher rating pointed toward AI rather than toward a human writer.

\begin{figure}[t]
\centering
\includegraphics[width=\linewidth]{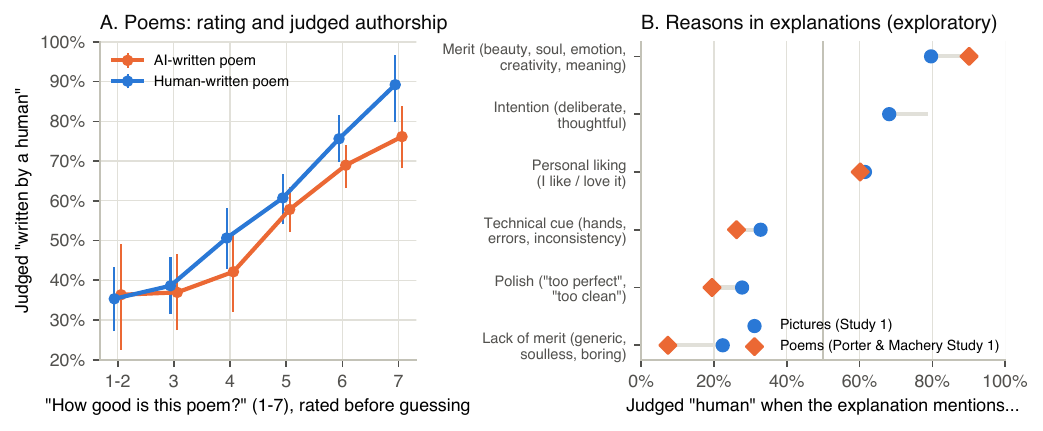}
\caption{(A) Share of poems judged ``written by a human'' by the rating given just before the judgment (Study~2, told-nothing condition), by true source, with 95\% intervals from 1{,}000 bootstrap resamples of participants. (B) Exploratory: share of judgments ``human'' among explanations matched by each keyword category, for pictures (Study~1) and poems (Study~1 of \citeauthor{porter2024ai}). Intention is shown for pictures only; fewer than 30 poem explanations matched it.}
\label{fig:poems}
\end{figure}

Higher-rated poems were nevertheless judged human more often (Table~\ref{tab:poetry} and Figure~\ref{fig:poems}A). Each additional point of rated quality went with an \PSlope{}~pp higher probability of a ``human'' judgment (SE \PSlopeSE{}), holding participant and poem fixed. The slope was the same for AI-written poems (\PSlopeAI{}~pp) and human-written ones (\PSlopeH{}~pp; difference \PSlopeContrast{}~pp, 95\% CI \PSlopeContrastLo{} to \PSlopeContrastHi{}), and leaving out one poem at a time moves it only between \PLooSlopeMin{} and \PLooSlopeMax{}~pp. Every one of the fourteen rated qualities was positively associated with a ``human'' judgment, from \PQualMin{}~pp per standard deviation (wit) to \PQualMax{}~pp (Appendix Table~\ref{tab:qualities}). The association was no weaker among participants who write poetry (\PBgWrite{}~pp per point, SE \PBgWriteSE{}).

The poem or poems a participant rated highest were judged human-written \PTopCalledH{} of the time, against \PNonTopCalledH{} for the same participant's other poems. With fixed effects, the top-rated set was \PTop{}~pp (SE \PTopSE{}) more likely to be judged human; among the \PTopUniqueN{} participants with a single top-rated poem, the estimate was \PTopUnique{}~pp (SE \PTopUniqueSE{}). Unlike the rating slope, this top-rated contrast differed by source: \PTopAI{}~pp for AI-written and \PTopH{}~pp for human-written poems (difference \PTopContrast{}~pp, 95\% CI \PTopContrastLo{} to \PTopContrastHi{}).

\begin{table}[t]
\centering
\caption{Liking and judged authorship of poems (Study~2). Linear models with participant and poem fixed effects in the told-nothing condition (\PNPeople{} participants, \PNJudg{} judgments); standard errors clustered by participant. $^a$The label effect compares the two framing conditions (\PNLabel{} participants) with poem fixed effects; the human label named a poet, the AI label ``an AI program''.}
\label{tab:poetry}
\small
\begin{tabular}{lrr}
\toprule
Quantity & Estimate (SE) & Unit \\
\midrule
\multicolumn{3}{l}{\emph{Told-nothing condition: judged ``written by a human''}} \\
\quad Each point on ``How good is this poem?'' (1--7) & $+$11.5 (0.9) & pp \\
\quad \quad AI-written poems & $+$11.7 (1.4) & pp \\
\quad \quad Human-written poems & $+$11.4 (1.1) & pp \\
\quad \quad Human minus AI & $-$0.3 (1.5) & pp \\
\quad Top-rated poem(s) vs.\ the respondent's other poems & $+$23.6 (2.8) & pp \\
\quad \quad AI-written poems & $+$18.5 (3.2) & pp \\
\quad \quad Human-written poems & $+$32.7 (4.4) & pp \\
\quad \quad Human minus AI & $+$14.3 (5.0) & pp \\
\quad Participants with a single top-rated poem (n = 77) & $+$24.6 (5.0) & pp \\
\multicolumn{3}{l}{\emph{Rating of ``How good is this poem?''}} \\
\quad AI-written minus human-written (told nothing) & 1.12 (0.09) & points \\
\quad Believed human minus believed AI (told nothing, fixed effects) & 0.89 (0.08) & points \\
\quad Told human minus told AI (framing conditions, poem fixed effects)$^a$ & 0.81 (0.09) & points \\
\quad Difference between the two gaps & 0.08 (0.12) & points \\
\bottomrule
\end{tabular}

\end{table}

The two framing conditions place this association on a familiar scale. Telling participants that the poems were written by named poets rather than by an AI program raised their ratings of how good the poems were by \PLabel{} points (SE \PLabelSE{}). In the told-nothing condition, poems that a participant judged human were rated \PGap{} points higher (SE \PGapSE{}) than poems the same participant judged AI-written, holding the poem fixed. The difference between the two is \PGapMinusLabel{} points (95\% CI \PGapMinusLabelLo{} to \PGapMinusLabelHi{}). The average of all fourteen ratings gives a similar comparison (\PGapComp{} against \PLabelComp{} points). The rating gap that accompanies people's own source judgments is thus of the same order as the effect of an explicit authorship label.

\subsection{Reasons in explanations (RQ5, exploratory)}

Finally, we turn to the reasons people gave. The merit keywords matched few explanations, and the explanations they matched usually accompanied a ``human'' judgment (Table~\ref{tab:reasons} and Figure~\ref{fig:poems}B). Explanations of picture judgments that matched the merit keywords, such as ``it has a gorgeous composition'' or ``seems to have artistic vision'', accompanied a ``human'' judgment \TImgMerit{} of the time. Explanations that matched the lack-of-merit keywords, such as ``generic'' or ``soulless'', accompanied a ``human'' judgment \TImgNoMerit{} of the time. The corresponding shares for poems were \TPoemMerit{} and \TPoemNoMerit{}. Matches for intention also accompanied ``human'' judgments (\TImgIntent{}). Matches for technical cues accompanied ``AI'' judgments (\TImgFlaw{} human), as did matches for polish (\TImgPolish{}). Within the same picture or poem, these differences remain (Table~\ref{tab:reasons}).

\begin{table}[t]
\centering
\caption{Reasons in free-text explanations (exploratory). Share of explanations matched by each keyword category (categories are not exclusive), share of the corresponding judgments that were ``human'', and the difference in that share between matched and unmatched explanations, within the same picture or poem (pp, 95\% CI clustered by respondent). Pictures: six pictures of Study~1. Poems: optional explanations in Study~1 of \citet{porter2024ai}. $^a$Fewer than 30 explanations.}
\label{tab:reasons}
\scriptsize
\setlength{\tabcolsep}{3.5pt}
\begin{tabular}{lrrrrrr}
\toprule
 & \multicolumn{3}{c}{Pictures (Study 1)} & \multicolumn{3}{c}{Poems (P\&M Study 1)} \\
\cmidrule(lr){2-4}\cmidrule(lr){5-7}
Explanation mentions\ldots & Share & Human & Diff.\ (95\% CI) & Share & Human & Diff.\ (95\% CI) \\
\midrule
Merit: beauty, soul, emotion, meaning, creativity & 1.7\% & 80\% & $+$20 [$+$17, $+$22] & 8.5\% & 90\% & $+$40 [$+$36, $+$43] \\
Intention: deliberate, thoughtful & 2.9\% & 68\% & $+$11 [$+$9, $+$14] & 0.4\% & 78\%$^a$ & $+$27 [$+$11, $+$42] \\
Personal liking: I like / love it & 3.3\% & 61\% & $+$7 [$+$5, $+$10] & 4.2\% & 60\% & $+$7 [$+$1, $+$14] \\
Technical cue: hands, errors, inconsistency & 13.9\% & 33\% & $-$19 [$-$20, $-$18] & 3.2\% & 26\% & $-$26 [$-$33, $-$18] \\
Polish: too perfect, too clean, too smooth & 1.4\% & 28\% & $-$19 [$-$22, $-$15] & 0.7\% & 19\% & $-$36 [$-$49, $-$23] \\
Lack of merit: generic, soulless, boring, clich\'e & 2.8\% & 22\% & $-$23 [$-$25, $-$21] & 4.4\% & 7\% & $-$50 [$-$54, $-$46] \\
\midrule
All explanations (n = 49{,}268; 5{,}238) & 100\% & 50\% & & 100\% & 53\% & \\
\bottomrule
\end{tabular}

\end{table}

The merit keywords matched \TImgMeritShare{} of picture explanations and \TPoemMeritShare{} of poem explanations, whereas the technical-cue keywords matched \TImgFlawShare{} of picture explanations. Fans' explanations of their own favourite matched the merit keywords in \TFansMerit{} of cases, against \TOthersMerit{} for other respondents explaining the same pictures. Some explanations thus state the beauty-is-human inference in so many words, and fans described their favourite's merit more often than other respondents did.

\section{Discussion}
\label{sec:discussion}

People who could not reliably tell AI-made art from human art judged the art they liked best to be human-made far more often than other people judged the same works. Across 50 pictures and \NFav{} respondents, the favourite was \Main{}~pp more likely to be judged human-made than the respondent's general tendency and the picture's general appearance predicted. The association was positive for every picture and similar for AI-made and human-made favourites. It appeared under every reported strategy, at every level of familiarity with human art, and at every level of opinion of AI art. It also appeared in a second medium with ratings recorded before each source judgment. In both media, liking was not positively diagnostic of human authorship: AI-made pictures were slightly overrepresented among favourites, and AI-written poems were rated higher than human-written ones.

\subsection{What the association adds to prior evidence}

Laboratory studies had shown that works people believe to be machine-made receive lower ratings, in images, music, haiku, and poems \citep{chamberlain2018putting,shank2023ai,hitsuwari2023does,grassini2024understanding,porter2024ai}. Our results add two things to this evidence. First, they characterize the association work by work. With the same 50 pictures judged by thousands of people, the association could be estimated for each picture: it was positive for all of them, ranging from \PicMin{} to \PicMax{}~pp, and it was similar for AI-made and human-made works. It is therefore not confined to correctly recognized human works. Second, they connect the association to people's stated opinion of AI art and to the source of the works they chose, where earlier work examined personality and attitudes toward technology \citep{grassini2024understanding}. People who disliked AI art chose fewer AI-made favourites but were more likely to see the ones they chose as human-made. Part of that difference reflected which AI-made pictures they chose, and part reflected how they judged the same pictures. The poetry reanalysis shows that the pattern holds in another medium when ratings come first, which extends the original authors' analysis of the same data \citep{porter2024ai}.

\subsection{Explaining the association}

Section~\ref{sec:accounts} described four processes that could link liking and believed authorship. The evidence bears on each, but none is excluded.

\emph{Belief-driven evaluation} is supported by the label benchmark: in the poetry experiment, an authorship label moved ratings by about as much as the rating gap that accompanied people's own source judgments. It is also supported by the attitude gradient: the association was strongest among people most negative about AI art. In Study~1, where the favourite was named after all source judgments, people could have chosen among the pictures they had already judged human.

\emph{Aesthetic response as evidence}, the beauty-is-human inference, is consistent with three observations. The association remained substantial among people who rate AI art most highly. It appeared when ratings were recorded before each source judgment. And explanations that matched the merit keywords usually accompanied a ``human'' judgment. None of these observations is decisive. People who rate AI art highly may still value human authorship, the survey did not measure that preference, and reporting a rating first does not mean that the rating was formed before any belief about the source.

The results on recognition and source narrow the room for \emph{shared perceptions of the work}. The association was as large among respondents who had seen none of the pictures before, and similar for AI-made and human-made works. A perception that only real human work evokes cannot produce it. A perception that AI-made and human-made works can both evoke, such as a sense of intention or meaning, remains a candidate, and so does a cue that is more common in human work but also appears in AI work. Respondent and picture fixed effects leave such person-specific perceptions of particular pictures unmeasured. Viewers do report emotions and perceive intention in computer-generated art \citep{demmer2023does}.

\emph{Response consistency} predicts the association in either order of reporting, as Chamberlain et al.'s order manipulation found \citep{chamberlain2018putting} and as our two studies found. It cannot be excluded with these data.

These processes are not mutually exclusive, and the first two may reinforce each other: a belief that a person made a work can raise its appeal, and its appeal can raise the belief that a person made it. Separating them requires experiments that hold the work fixed while changing either the experience or the belief. One discriminating prediction follows from research on feelings as information. If people use their pleasure as evidence of a human maker, the association should weaken when an incidental source of the pleasure is made salient, as misattributed moods lose their influence once their cause is pointed out \citep{schwarz1983mood}. Belief-driven evaluation predicts little change. The contrast holds only if the reminder does not itself change source beliefs, attention, response consistency, or how people read the work, so the experiment would need an active control and measures of both enjoyment and source belief.

\subsection{Opinions, choices, and beliefs}

An opinion of AI art is a judgment about a category, and it does not commit a person to disliking every AI-made work. Our results show how such opinions relate to the specific works people choose. People who disliked AI art chose AI-made favourites less often. When they did choose one, it tended to be a picture that other people also took to be human, and they themselves were more likely to judge that same picture human. As a result, the share of people who named an AI-made favourite they believed a person had made stayed between \JointMin{} and \JointMax{} across the opinion groups. The pattern does not tell us which way the influences run. Opinions may shape which works people choose and how they read them, or experiences with particular works may shape the opinions people report afterwards. It does show that a negative opinion of AI art neither keeps AI-made work out of people's favourites nor keeps them from believing that work to be human-made. This qualifies how aversion to AI art is described. Laboratory studies measure aversion as lower ratings for work labelled as AI-made \citep{millet2023defending,chiarella2022investigating}. Without labels, the people who report the strongest aversion still chose an AI-made favourite almost half the time (\AttFavAIOne{}, against \ShareImgAI{} of the pictures being AI-made), and they mostly believed it to be human-made.

\subsection{Practical implications}

The findings suggest hypotheses for practice that can be tested directly.

\paragraph{Disclosure.} Rules that require AI-generated content to be labelled, such as the transparency obligations of the European Union's AI Act \citep{euaiact2024}, are usually justified by the need to prevent deception. Our results point to a further consequence. Without labels, the works people like tend to be the works they credit to people, whatever their source. Disclosing AI authorship after an author-blind first impression lowers evaluations \citep{popescu2026code}. A natural next test is whether reliable disclosure changes choices, sharing, or willingness to pay most for works that were liked and believed to be human. Our data contain no purchases or payments, so any consequence for markets in human-made work is a hypothesis for such studies.

\paragraph{Detection advice.} Common advice for spotting AI images directs attention to technical defects such as hands, garbled text, and inconsistent lighting. Looking for such errors was the strategy our respondents reported most often, and those respondents showed the favourite association as strongly as anyone. A reported main strategy does not measure training or exclusive reliance on that strategy, and the test removed obvious defects from its AI-made pictures, which makes it a hard setting for error-based strategies. Whether instruction that warns people about the pull of their own liking improves their judgments of the works they like most is a question for a randomized comparison with error-based advice.

\paragraph{Research design.} Studies that ask people both to rate a work and to judge its source should expect the two answers to be linked whichever comes first. Reading the link as the effect of believed authorship on liking, or as the effect of liking on believed authorship, requires a design that manipulates one of them.

\subsection{Limitations and future research}

Several features of the data define directions for further work.

\paragraph{Population and works.} The respondents of Study~1 were readers of one blog who chose to take the test. They were likely more interested in AI than the general population. The 50 pictures were curated to be difficult, and estimates refer to these pictures and, in Study~2, to ten poems. The picture-level estimates and the poetry reanalysis describe these sets of works; they do not predict the association for works selected differently, or for favourites chosen from a different set. The public file does not identify respondents, so duplicate participation cannot be checked, and some respondents may have seen discussions of the test beforehand. Representative samples of people, and of works drawn from what people encounter online, would establish how widespread the association is.

\paragraph{Direction.} Neither study manipulates liking. Study~2 records ratings before each source judgment, and its framing conditions show that an authorship label changes ratings, but both judgments in the told-nothing condition are the participant's own. The most informative next study would randomize an evaluative experience while holding the work fixed, measure source beliefs on a continuous scale, and include a condition that makes the incidental source of the experience salient. Such an intervention could also act on source beliefs through familiarity, mood, or attention, so manipulation checks and an active control would be needed.

\paragraph{Measures.} Opinions of AI art were reported after the test, and the survey did not separate views about quality from preferences for human authorship or ethical concerns; measures taken before exposure would sharpen the comparison between stated opinion and choice. Familiarity, strategy, and expertise were self-reported in single items, and the weaker association among the \FamAIExpertN{} respondents who described AI art as their job or greatest passion is a lead for a better-powered study. The signal detection summaries aggregate across respondents with different favourites, and a hierarchical model of trial-level judgments with person and item variation \citep{rouder2005introduction} could estimate criterion and sensitivity within persons. The keyword measure of reasons has not been validated against human coding, and it counts only explicit mentions in brief answers. A coding scheme applied by independent coders, which separates the quality a statement refers to from the direction of the source judgment, would turn the exploratory description into a measurement and could estimate how often people reason from merit without naming it.

\paragraph{Consequences.} The data record judgments and ratings, not choices with costs. Studies that measure purchases, sharing, or credit given to creators, before and after reliable disclosure, would show whether the association affects behaviour beyond the survey.

\section{Conclusion}
\label{sec:conclusion}

When people cannot tell who made a work of art, the works they like most are the ones they judge to be made by people. In a large online test, this association held for every picture, for AI-made and human-made pictures alike, under every reported detection strategy, and at every level of opinion of AI art. A reanalysis of a poetry experiment found it again when ratings came first. People who disliked AI art chose fewer AI-made favourites but believed more of them to be human-made, so that roughly a third to two fifths of respondents in every opinion group favoured an AI-made picture they took to be human. As generated pictures, poems, and music become indistinguishable from human work, what people believe about who made a work will be bound up with what they like. These findings motivate reliable provenance information rather than reliance on subjective liking as evidence of authorship.

\bibliographystyle{unsrtnat}
\bibliography{references}

\appendix
\section{Sample flow in Study 1}
\label{app:flow}

\begin{table}[H]
\centering
\caption{Respondents included in and excluded from the favourite analysis (Study~1). Included respondents named a favourite and judged it. Shares are of all respondents in each group.}
\label{tab:flow}
\small
\begin{tabular}{lrr}
\toprule
 & Included & Excluded \\
\midrule
Respondents & 9{,}152 & 1{,}971 \\
Pictures judged (of 50) & 49.6 & 44.8 \\
Accuracy & 60.9\% & 59.5\% \\
Share of ``human'' answers & 51.7\% & 52.0\% \\
Opinion of AI art (1--5) & 3.12 & 2.99 \\
Opinion of AI art missing & 1\% & 42\% \\
Art is job or greatest passion & 4.1\% & 1.9\% \\
Familiarity with AI art (0--3) & 0.97 & 0.91 \\
Familiarity questions missing & 0\% & 42\% \\
\bottomrule
\end{tabular}

\end{table}

\section{The pictures of Study 1}
\label{app:pictures}

Table~\ref{tab:pictures} lists the 50 pictures in the order of the test. For each picture it gives the true source, the share of all respondents who judged it human-made, and the number of respondents who named it as their favourite. It then gives two measures of the favourite association: the raw gap, which is the share of fans who judged the picture human minus the share among all other respondents in the analysis sample, and the picture-specific estimate with respondent and picture fixed effects. Gaps for pictures with fewer than 30 fans are in parentheses. Both the raw gap and the adjusted estimate are positive for all 50 pictures.

\paragraph{Co-judgment clusters.} For each respondent, we subtracted the respondent's mean ``human'' rate from each of their judgments, computed the correlation between pictures across respondents, and grouped pictures by average-linkage hierarchical clustering on one minus the correlation. Pictures in the same cluster are the ones that the same people tend to call human, beyond their general tendency to do so; the clusters reflect style and any other shared appearance. With four clusters, the groups contain 18, 17, 11, and 4 pictures. The cluster numbers in Table~\ref{tab:pictures} refer to the four-cluster solution; the six- and eight-cluster solutions give the estimates in Table~\ref{tab:specs}. For the held-out check, respondents were split at random into two halves (\HeldN{} respondents in the estimation half), clusters were built from one half, and the favourite association was estimated in the other.

{\small
\setlength{\tabcolsep}{3.5pt}
\begin{longtable}{rlcrrrrrrc}
\caption{The 50 pictures of Study 1. Source is the true origin; ``Judged human'' is the share of all respondents calling it human; fans are respondents who named it their favourite and also judged it (Fav.: number of respondents naming it). Gap is the raw difference between fans and other respondents; Adjusted is the picture-specific favourite association with respondent and picture fixed effects (SE clustered by respondent). Cl.\ is the four-cluster co-judgment solution (Appendix~\ref{app:pictures}).}\label{tab:pictures}\\
\toprule
No. & Picture & Source & Judged human & Fav. & Fans: human & Others: human & Gap & Adjusted (SE) & Cl. \\
\midrule
\endfirsthead
\toprule
No. & Picture & Source & Judged human & Fav. & Fans: human & Others: human & Gap & Adjusted (SE) & Cl. \\
\midrule
\endhead
\bottomrule
\endfoot
1 & Angel Woman & Human & 46\% & 221 & 80\% & 45\% & $+$35.2 & $+$35.2 (2.8) & 4 \\
2 & Saint In Mountains & Human & 78\% & 329 & 89\% & 77\% & $+$12.1 & $+$12.9 (1.8) & 2 \\
3 & Blue Hair Anime Girl & Human & 51\% & 67 & 71\% & 51\% & $+$19.9 & $+$17.4 (6.2) & 4 \\
4 & Girl In Field & AI & 38\% & 42 & 60\% & 38\% & $+$22.0 & $+$21.1 (7.4) & 1 \\
5 & Double Starship & Human & 56\% & 146 & 79\% & 57\% & $+$21.9 & $+$21.9 (3.5) & 4 \\
6 & Bright Jumble Woman & AI & 52\% & 63 & 70\% & 51\% & $+$18.7 & $+$19.9 (5.7) & 3 \\
7 & Cherub & AI & 56\% & 43 & 86\% & 57\% & $+$29.5 & $+$30.6 (5.6) & 1 \\
8 & Praying In Garden & Human & 67\% & 167 & 95\% & 66\% & $+$28.2 & $+$27.6 (2.0) & 2 \\
9 & Tropical Garden & Human & 80\% & 161 & 96\% & 80\% & $+$16.7 & $+$17.8 (1.8) & 2 \\
10 & Ancient Gate & AI & 9\% & 327 & 22\% & 9\% & $+$13.8 & $+$13.6 (2.4) & 4 \\
11 & Green Hills & AI & 74\% & 385 & 90\% & 73\% & $+$17.3 & $+$18.0 (1.6) & 1 \\
12 & Bucolic Scene & Human & 85\% & 346 & 95\% & 85\% & $+$10.1 & $+$12.2 (1.3) & 2 \\
13 & Anime Girl In Black & AI & 20\% & 132 & 30\% & 19\% & $+$10.3 & $+$9.8 (4.1) & 4 \\
14 & Fancy Car & Human & 23\% & 23 & 48\% & 23\% & ($+$24.5) & $+$21.3 (10.5) & 4 \\
15 & Greek Temple & Human & 40\% & 90 & 70\% & 40\% & $+$29.6 & $+$30.1 (5.0) & 2 \\
16 & String Doll & AI & 10\% & 280 & 29\% & 9\% & $+$20.2 & $+$21.0 (2.8) & 4 \\
17 & Angry Crosses & AI & 81\% & 94 & 96\% & 80\% & $+$15.5 & $+$13.9 (2.3) & 3 \\
18 & Rainbow Girl & Human & 50\% & 134 & 81\% & 51\% & $+$29.9 & $+$29.9 (3.6) & 4 \\
19 & Creepy Skull & Human & 86\% & 159 & 96\% & 86\% & $+$10.0 & $+$10.2 (1.7) & 2 \\
20 & Leafy Lane & AI & 66\% & 549 & 85\% & 65\% & $+$20.7 & $+$20.1 (1.6) & 1 \\
21 & Ice Princess & AI & 15\% & 116 & 28\% & 15\% & $+$12.2 & $+$10.0 (4.4) & 4 \\
22 & Celestial Display & Human & 52\% & 473 & 83\% & 52\% & $+$31.5 & $+$32.9 (1.8) & 4 \\
23 & Mother And Child & AI & 39\% & 211 & 76\% & 38\% & $+$37.6 & $+$36.7 (3.0) & 1 \\
24 & Fractured Lady & AI & 52\% & 88 & 89\% & 52\% & $+$36.9 & $+$35.2 (3.5) & 3 \\
25 & Giant Ship & Human & 16\% & 178 & 38\% & 16\% & $+$21.2 & $+$22.3 (3.7) & 4 \\
26 & Muscular Man & AI & 32\% & 65 & 38\% & 31\% & $+$7.4 & $+$5.3 (6.1) & 1 \\
27 & Minaret Boat & AI & 17\% & 131 & 43\% & 17\% & $+$25.5 & $+$25.3 (4.4) & 4 \\
28 & Purple Squares & Human & 79\% & 116 & 94\% & 79\% & $+$15.1 & $+$14.4 (2.4) & 2 \\
29 & People Sitting & Human & 79\% & 152 & 95\% & 78\% & $+$17.4 & $+$17.7 (1.8) & 2 \\
30 & Girl In White & Human & 52\% & 133 & 89\% & 52\% & $+$37.7 & $+$37.4 (2.8) & 2 \\
31 & Riverside Cafe & AI & 69\% & 642 & 88\% & 67\% & $+$20.3 & $+$19.4 (1.4) & 1 \\
32 & Serene River & Human & 74\% & 151 & 95\% & 73\% & $+$22.6 & $+$22.8 (1.8) & 2 \\
33 & Turtle House & AI & 21\% & 211 & 39\% & 21\% & $+$17.8 & $+$19.6 (3.5) & 4 \\
34 & Still Life & AI & 57\% & 122 & 88\% & 56\% & $+$32.6 & $+$31.7 (3.0) & 1 \\
35 & Wounded Christ & Human & 72\% & 115 & 90\% & 72\% & $+$17.6 & $+$16.1 (3.0) & 2 \\
36 & White Blob & Human & 74\% & 71 & 97\% & 74\% & $+$22.8 & $+$24.8 (2.2) & 2 \\
37 & Weird Bird & AI & 23\% & 37 & 39\% & 22\% & $+$16.6 & $+$15.1 (8.1) & 4 \\
38 & Ominous Ruin & AI & 31\% & 191 & 64\% & 29\% & $+$35.0 & $+$33.8 (3.4) & 1 \\
39 & Vague Figures & Human & 79\% & 65 & 91\% & 80\% & $+$11.2 & $+$12.5 (3.7) & 2 \\
40 & Dragon Lady & AI & 20\% & 219 & 38\% & 19\% & $+$18.9 & $+$18.6 (3.4) & 4 \\
41 & White Flag & Human & 68\% & 384 & 93\% & 67\% & $+$26.3 & $+$26.4 (1.4) & 2 \\
42 & Woman Unicorn & Human & 75\% & 113 & 94\% & 75\% & $+$18.6 & $+$18.3 (2.5) & 2 \\
43 & Rooftops & AI & 30\% & 237 & 66\% & 28\% & $+$37.7 & $+$37.8 (3.1) & 4 \\
44 & Paris Scene & AI & 84\% & 336 & 96\% & 83\% & $+$12.7 & $+$12.0 (1.2) & 1 \\
45 & Pretty Lake & AI & 55\% & 69 & 88\% & 54\% & $+$34.5 & $+$32.2 (4.0) & 1 \\
46 & Landing Craft & AI & 24\% & 81 & 57\% & 23\% & $+$33.8 & $+$33.4 (5.6) & 4 \\
47 & Flailing Limbs & Human & 79\% & 233 & 95\% & 78\% & $+$16.5 & $+$17.2 (1.5) & 2 \\
48 & Colorful Town & Human & 49\% & 41 & 80\% & 47\% & $+$33.0 & $+$30.9 (6.2) & 2 \\
49 & Mediterranean Town & AI & 65\% & 359 & 89\% & 64\% & $+$25.5 & $+$25.1 (1.8) & 4 \\
50 & Punk Robot & AI & 40\% & 74 & 77\% & 39\% & $+$37.3 & $+$38.1 (4.9) & 3 \\
\end{longtable}

}

\section{Poetry: all fourteen rated qualities}
\label{app:qualities}

\begin{table}[H]
\centering
\caption{Judging a poem human-written by each rated quality (Study~2, told-nothing condition). Each row is a separate linear model of the ``human'' judgment on the standardized rating, with participant and poem fixed effects; standard errors clustered by participant. The qualities are correlated measures of one evaluation.}
\label{tab:qualities}
\small
\begin{tabular}{lr}
\toprule
Rated quality (1--7) & Per SD, pp (SE) \\
\midrule
good (overall quality) & $+$18.7 (1.5) \\
meaningful & $+$18.0 (1.6) \\
beautiful & $+$17.8 (1.6) \\
conveys a mood or emotion & $+$17.7 (1.5) \\
imagery & $+$16.9 (1.5) \\
moving & $+$16.6 (1.6) \\
inspiring & $+$16.6 (1.6) \\
sound & $+$16.3 (1.6) \\
original & $+$16.0 (1.5) \\
theme & $+$15.9 (1.6) \\
profound & $+$15.1 (1.6) \\
rhythm & $+$14.9 (1.7) \\
lyrical & $+$14.0 (1.6) \\
witty & $+$11.9 (1.8) \\
\bottomrule
\end{tabular}

\end{table}

\section{The keyword measure of reasons}
\label{app:text}

Explanations were lower-cased and matched on whole words, using the term lists in Table~\ref{tab:dictionary}, where an asterisk marks any ending. A merit term preceded within three words by a negation (no, not, lack, lacks, lacked, lacking, without, devoid, void, missing, nothing, zero, little, absent, absence, never, isn't, doesn't, don't, didn't, wasn't, aren't, can't, cannot, unclear) was counted as lack of merit instead of merit, and such a negated span did not prevent other, non-negated merit terms in the same explanation from counting as merit. Before technical cues were matched, we removed phrases that contain a cue word without referring to the work: ``on the one hand'' and ``on the other hand'', ``hand-drawn'', ``hand-painted'' and similar forms, ``by hand'' and ``first-hand'', ``human perspective'', ``personal perspective'' and similar forms, and ``from a \ldots\ perspective''.

\begin{table}[H]
\centering
\caption{Keyword categories used to code explanations. An asterisk marks any ending; categories are not exclusive.}
\label{tab:dictionary}
\small
\begin{tabular}{>{\raggedright\arraybackslash}p{0.17\linewidth}>{\raggedright\arraybackslash}p{0.76\linewidth}}
\toprule
Category & Terms \\
\midrule
Merit & beaut*, gorgeous, lovely, stunning, soul, soulful, emotion, emotions, emotional*, emotive, evocative, expressive, meaningful, meaning, profound, creativ*, imaginative, inspired, inspiring, passion*, charm*, personality, alive, lively, vision, mood, whimsical, playful, humanity, human touch, human experience, sincer*, authentic* \\
Intention & intention*, deliberate*, thoughtful*, purposeful* \\
Personal liking & I like / love / enjoy / adore (optionally ``really''), like it, love it, like this, love this, favourite, favorite, my fav \\
Technical cue & hand(s), finger(s), error(s), mistake(s), inconsisten*, incoheren*, nonsens*, doesn't make sense, makes no sense, no sense, artifact(s), artefact(s), glitch*, warp*, distort*, merg*, melt*, perspective, anatom*, proportion*, asymmetr*, physics, illogical, garbled, deformed, malformed \\
Polish & too perfect / polished / clean / smooth / good / neat / detailed / precise / crisp / sharp / symmetric* / pretty / glossy / shiny / refined / well done; sheen; glossy; overly polished / smooth / detailed / perfect \\
Lack of merit & soulless, lifeless, generic, bland, boring, cheesy, tacky, clich*, trite, meaningless, pointless, uninspir*, uninteresting, sterile, vapid, emotionless, doesn't say anything, says nothing, nothing to say, no point; negated merit terms \\
\bottomrule
\end{tabular}
\end{table}

The measure counts explicit mentions and misses paraphrases. It also misreads some explanations, and its errors need not be random with respect to the judgment. A respondent who writes that ``AI has no emotions'' as a reason for judging a poem human is coded as citing a lack of merit; a mention of hands need not assert a defect; and ``too perfect'' can support either judgment, depending on what the respondent believes AI can do. Table~\ref{tab:quotes} shows randomly drawn examples, including such misreadings.

{\small
\begin{longtable}{p{0.13\linewidth}p{0.08\linewidth}p{0.07\linewidth}p{0.58\linewidth}}
\caption{Verbatim explanations, drawn at random among those of 15--110 characters within each category (spelling as written).}\label{tab:quotes}\\
\toprule
Category & Source & Judged & Explanation \\
\midrule
\endfirsthead
\toprule
Category & Source & Judged & Explanation \\
\midrule
\endhead
\bottomrule
\endfoot
Merit & images & human & ``Use of positive and negative space, feels vaguely creative'' \\
Merit & images & human & ``i dont feel like ai would be this creative'' \\
Merit & images & AI & ``first impression was creatively-arranged cut paper, but then i noticed bizarre irregular silhouettes.'' \\
Merit & images & AI & ``Same structure as there was before, mess behind does have meaning'' \\
Lack of merit & images & human & ``Every small stroke has a meaning, making a shape or a light. No meaningless or unclear spots'' \\
Lack of merit & images & human & ``Looks too boring for an AI'' \\
Lack of merit & images & AI & ``Vague and generic'' \\
Lack of merit & images & AI & ``Chaotic soulless blob of graffiti. Soulless either way.'' \\
Intention & images & human & ``The way both negative and positive space are used feels more intentional than I'd expect for AI'' \\
Intention & images & human & ``The abstraction seems too purposeful to be an AI, but maybe I'm wrong!'' \\
Intention & images & AI & ``The lines seem too sharp yet random to be intentional'' \\
Intention & images & AI & ``I'd expect more clear meaning or intention from a human artist'' \\
Polish & images & human & ``No obvious errors, lacks glossy/fake antique effects'' \\
Polish & images & human & ``The symmetry of the figure is probably too good for AI, but I don't know.'' \\
Polish & images & AI & ``way too shiny, proportions are funky'' \\
Polish & images & AI & ``Brushstrokes look a little too precise'' \\
Technical cue & images & human & ``Small ''unnecessary'' details that are not artefacts'' \\
Technical cue & images & human & ``Has perspective mistakes that I don't think a machine would make'' \\
Technical cue & images & AI & ``It looks silly, dreamlike, and there are many details that seem nonsensically placed'' \\
Technical cue & images & AI & ``Too many colors in random places to be human. Looks like there are AI artifacts.'' \\
Merit & poems & human & ``beautiful prose.'' \\
Merit & poems & human & ``It was beautifully written.'' \\
Merit & poems & AI & ``very expressive'' \\
Merit & poems & AI & ``There or feelings and emotions potrayed in the poem'' \\
Lack of merit & poems & human & ``this peom has somewhat emotions in it wich sounds really touching and an AI has no emotions to go that far'' \\
Lack of merit & poems & human & ``AI is not that expressive'' \\
Lack of merit & poems & AI & ``It's kind of generic.'' \\
Lack of merit & poems & AI & ``My sense is AI, or a poor effort man, but probably AI. feels lifeless, empty, uncommitted'' \\
Intention & poems & human & ``It looks too thoughtful to be AI'' \\
Intention & poems & human & ``Seems more thoughtful than just ''what word comes next'''' \\
Intention & poems & AI & ``Sounds pretty, but regularly breaks scansion/rhyme scheme in ways that don't sound intentional'' \\
Intention & poems & AI & ``this is very organized and thoughtful and well structured. It seems manufactured by AI'' \\
Polish & poems & human & ``This seems too perfect to be AI'' \\
Polish & poems & human & ``This seems too good to be written by an artificial intelligence'' \\
Polish & poems & AI & ``This poem seems too perfect.'' \\
Polish & poems & AI & ``Seems too perfect for a human'' \\
Technical cue & poems & human & ``a few grammatical errors'' \\
Technical cue & poems & human & ``The odd imagery made me suspect it was a human since it almost nonsensical images seemed more humanlike.'' \\
Technical cue & poems & AI & ``Difficult to read with some spelling errors. Seems more AI.'' \\
Technical cue & poems & AI & ``Makes no sense to me.'' \\
\end{longtable}

}

\end{document}